\documentclass[aps,12pt]{article}   
\usepackage[dvips]{graphics}  
\usepackage[dvips]{graphicx}  
\usepackage{float}     
\usepackage{epsfig}  
\usepackage{tabularx,longtable}  
\usepackage{color}  
\usepackage{fancybox}  
\usepackage{multicol}  

\newcommand{\bi}{\begin{itemize}}  
\newcommand{\ei}{\end{itemize}}  
\newcommand{\be}{\begin{equation}}  
\newcommand{\ee}{\end{equation}}

\newcommand{\bea}{\begin{eqnarray}}  
\newcommand{\eea}{\end{eqnarray}}  
\newcommand{\beastar}{\begin{eqnarray*}}  
\newcommand{\eeastar}{\end{eqnarray*}}

\newcommand{\eq}[1]{(\ref{#1})}

\usepackage{setspace}

\begin{document}

\title{Force-induced misfolding in RNA}

\author{M. Manosas${}^{1}$, I. Junier${}^{2}$, F. Ritort${}^{3,*}$\\  
{\small Departament de F\'{i}sica Fonamental, Facultat de F\'{\i}sica, Universitat de Barcelona} \\  
{\small Diagonal 647, 08028 Barcelona, Spain}}  

\footnotetext[{1}]{Present address: Laboratoire de Physique Statistique,  
Ecole Normale Sup\'erieure, Unit\'e Mixte de Recherche 8550 associ\'ee  
au Centre National de la Recherche Scientifique et aux Universit\'es  
Paris VI et VII, 24 rue Lhomond, 75231 Paris, France.}  
\footnotetext[{2}]{Present address: Programme Epig\'enomique,   
Genopole\textregistered, Tour \'Evry 2, 523 Terrasses de l'Agora, 91034 Evry cedex, France.}  
\footnotetext[{3}]{Additional affiliation: CIBER-BBN, Networking Centre on Bioengineering,   
Biomaterials and Nanomedicine.}


\maketitle  

\abstract{\bf RNA folding is a kinetic process governed by the 
competition of a large number of structures stabilized by the transient 
formation of base pairs that may induce complex folding pathways and the 
formation of misfolded structures. Despite of its importance in modern 
biophysics, the current understanding of RNA folding kinetics is limited 
by the complex interplay between the weak base-pair interactions that 
stabilize the native structure and the disordering effect of thermal 
forces. The possibility of mechanically pulling individual molecules 
offers a new perspective to understand the folding of nucleic 
acids. Here we investigate the folding and misfolding mechanism in RNA 
secondary structures pulled by mechanical forces. We introduce a model 
based on the identification of the minimal set of structures that 
reproduce the patterns of force-extension curves obtained in 
single molecule experiments.  The model requires only two fitting 
parameters: the attempt frequency at the level of individual base pairs 
and a parameter associated to a free energy correction that 
accounts for the configurational entropy of an exponentially large 
number of neglected secondary structures. We apply the model to 
interpret results recently obtained in pulling experiments in the 
three-helix junction S15 RNA molecule (RNAS15). We show that RNAS15 undergoes 
force-induced misfolding where force favors the formation of a stable 
non-native hairpin.  The model reproduces the pattern of unfolding and 
refolding force-extension curves, the distribution of breakage forces and 
the misfolding probability obtained in the experiments.}

\vspace*{1cm}


\newpage  

\section{\label{intro}Introduction}

Like proteins, RNAs have enzymatic, regulatory and structural
functions that are crucial for the correct operation of cells
\cite{doudna,moore}. RNA molecules are found in single stranded form
and are designed to fold into specific three-dimensional
conformations, called native states. RNA folding is a kinetic process
mainly governed by the interactions between complementary bases which
can lead to the formation of both native and non-native domains.  As a
result, folding into states that are structurally different from the
native state, usually referred as {\it misfolding}, can occur
\cite{herschlag}. Misfolded RNAs are not functional and can be harmful
to organisms \cite{chen}, just as misfolded proteins (e.g. prions)
that are involved in several diseases \cite{dobson}. Folding of
biomolecules, such as RNA molecules and proteins, is therefore a
subject of great importance in modern biophysics.  Under which
conditions misfolding is prone to occur?  What are the structural
elements that prevent folding into the native structure? Is it
possible to control misfolding by designing specific molecular
sequences?. To answer such questions modeling of biomolecular folding
is of great help. The competition between a very large number of
structures, that may lead to misfolding, makes modeling of folding a
difficult and challenging problem in biological physics where disorder
and frustration play a crucial role \cite{bundschuh,onuchic}. RNA
mostly folds in a hierarchical fashion dominated by the formation of
secondary structures \cite{brion,Tin1,zarrinkar,wu}. In contrast to
proteins where native state prediction is very difficult, it is
possible to infer the correct secondary structure of RNA molecules
from computer calculations (Mfold).
This makes RNA folding a more tractable theoretical problem than
protein folding. Bi-stability and misfolding in nucleic acids have
been recently investigated in temperature ramping \cite{ViaMelIsa06}
and force pulling \cite{LiBusTin07} experiments.

In this work we address the problem of folding/misfolding in RNA
molecules that are stretched by mechanical forces. Using single molecule
techniques it is nowadays possible to pull on individual molecules such
as biopolymers (e.g. nucleic acids, proteins, sugars...), molecular
complexes (e.g. motor proteins and DNA/protein fibers) or even to stretch
cells. Single molecule techniques provide valuable information about the
thermodynamics and kinetics of biomolecular processes, thereby enlarging our
knowledge of fundamental processes at the molecular and cellular
level \cite{Ritort06}. Among the most successful techniques in the field
are optical tweezers, AFM and magnetic tweezers, all them capable of
exerting forces in the piconewton (pN) range (1pN=$10^{-12}$N). Various studies have
investigated the unfolding/refolding of individual RNA molecules using optical tweezers. RNA
hairpins are typically unzipped at forces around 15pN where base pairs are
disrupted by the direct action of force. Folding kinetics in force is
of current interest as it provides an alternative route to investigate
the problem of molecular folding, complementary to studies of folding
by varying temperature or denaturant concentration. What is the main
effect of force in RNA folding? Under the action of
mechanical forces, the formation of
secondary contacts in RNA between bases located at distant segments of the
molecule is hampered by the stretching effect of the force. Starting
from a stretched state and by progressively decreasing the force,
folding is partially a sequential process in contrast to the
non-sequential mechanism observed in thermal folding \cite{Onoa}. Here we
introduce a phenomenological model, based on a sequential dynamics at
the level of individual base pairs, that is useful to investigate
folding and misfolding of RNA molecules that lack tertiary contacts.
We apply it to interpret and reproduce experimental results
recently obtained in the three-helix junction S15 RNA molecule,
hereafter referred as RNAS15, pulled by optical tweezers \cite{Coll}
(see Fig. \ref{f1}). These experiments consist of repeated force
cycles that start from the fully stretched molecule at high
forces. The force is first decreased down to low values to let the
molecule refold.  Next, it is increased up to the initial value in
order to unfold the molecule again \cite{Lip1}. In this way the
folding reaction can be monitored as a function of time. In such
experimental conditions, we show that RNAS15 undergoes force-induced
misfolding behavior as a consequence of the competition between the
formation of two hairpins that cannot coexist in the same
conformation. The computed misfolding probability, defined as the
probability to end up in the misfolded state at the end of the
relaxing process, is in good agreement with that obtained in the
experiments. We are also able to reproduce the experimental unfolding
and refolding force-extension trajectories, and obtain distributions
of breakage forces (i.e. the force at which the native structure
unfolds) that match the experimental ones at different loading rates.

\section{\label{fecs}Two unfolding patterns in RNAS15}

\begin{figure}[t] 
\centerline{\includegraphics[scale=0.5]{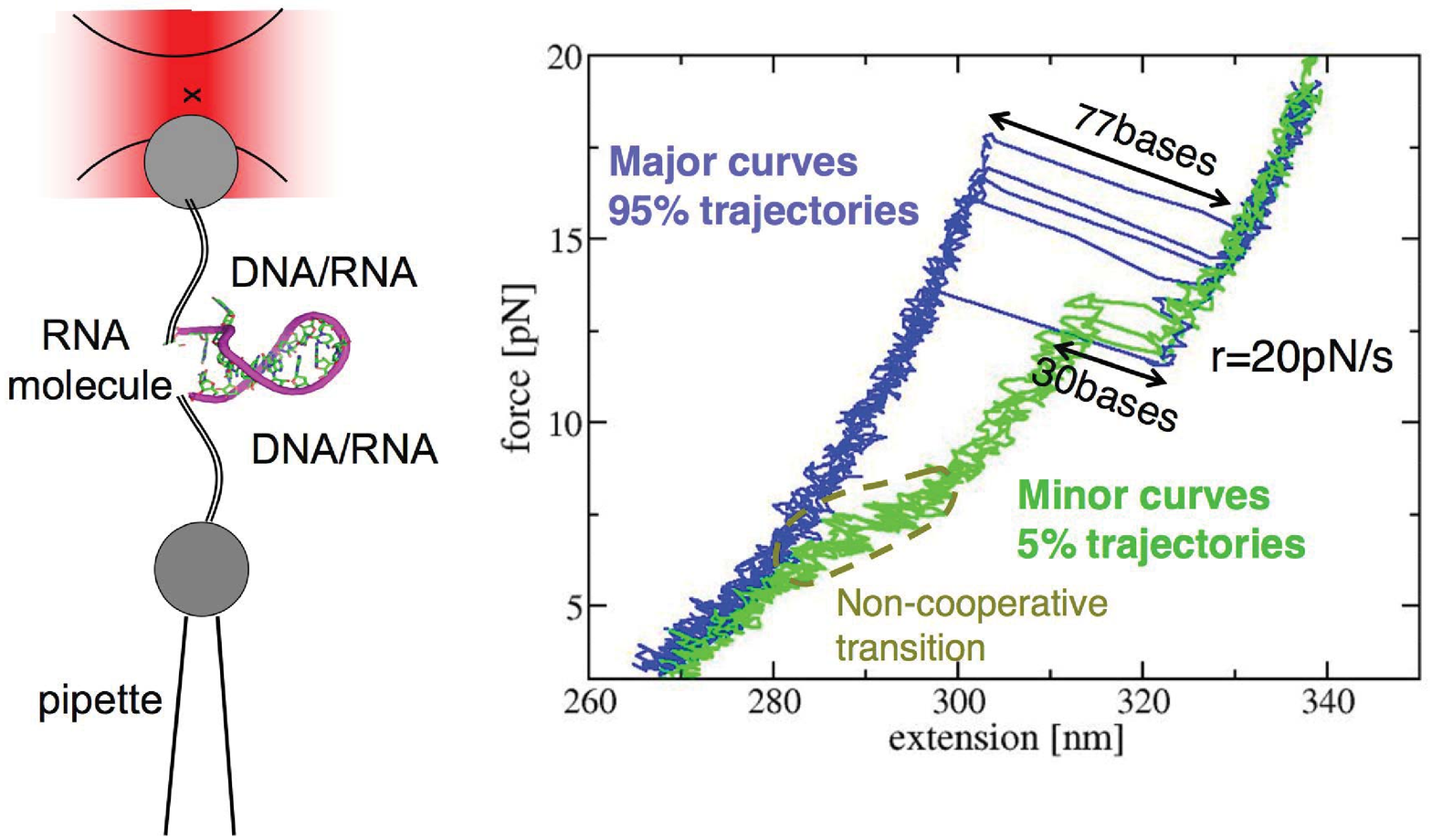}} 
\caption{\label{f1}}
\end{figure}

The present work is based on previous pulling experiments  
\cite{Coll} where optical tweezers \cite{Smith1} were used to induce  
unfolding and refolding in RNAS15 at room temperature ($T=298K$) in a  
solvent free of magnesium ions to avoid the formation of tertiary  
contacts. In these experiments a molecular construct is  
synthesized where the molecule RNAS15 is inserted between molecular  
DNA/RNA hybrid handles that provide enough space between the two beads  
to avoid non-specific interactions between the molecule and the beads,  
see Fig. \ref{f1}. The force applied on the molecular construct (RNAS15  
plus handles) is then ramped at constant speed  
\cite{Foot} between 2 pN and 20 pN at two loading rates, $r=12{\rm  
pN.s^{-1}}$ and $r=20{\rm pN.s^{-1}}$.  At 2 pN (20 pN), the  
thermodynamically stable state is the folded (stretched) state.  The  
output of the experiments is the force-extension curve that  
gives the force applied to the molecule as a function of the molecular  
extension.  During the unfolding part of the cycle (2 pN $\to$ 20 pN),  
two types of unfolding curves, referred to as {\it major} and {\it  
minor}, are observed (see Fig. \ref{f1}). The major curves correspond  
to approximately 95\% (90\%) of the trajectories at $r \simeq 12{\rm  
pN.s^{-1}}$ ($\simeq 20{\rm pN.s^{-1}}$).  The minor curves correspond  
to the rest $\simeq 5\%$ ($\simeq 10\%$).    

The major curves show a  
cooperative transition similar to that observed in the  
unzipping of small RNA hairpins \cite{Coll,Lip1}. Up to forces $\sim  
15$ pN, the force-extension curve corresponds to the stretching of the molecular  
handles used to manipulate the molecule \cite{Coll,Lip1}. The sudden  
large gain in the extension at forces around $15$ pN is consistent  
with the whole opening of RNAS15 that is $77$ bases long.  On the other  
hand, the minor curves do not show the typical stretching behavior of  
the handles at low forces ($f<14$ pN). In particular, a  
non-cooperative transition occurs at force values between 6 and 9  
pN. At these forces, the minor trajectories show large fluctuations in  
the extension (Fig. \ref{f1}) suggesting the presence of fast  
conformational events where the molecule partially unfolds and  
refolds. Moreover, the cooperative transition observed in the minor  
curves at forces around $14$ pN corresponds to the opening of a $\sim  
30$ bases domain that is much shorter than the total length of  
the RNA molecule.  

\begin{figure}[t] 
\centerline{\includegraphics[scale=0.5]{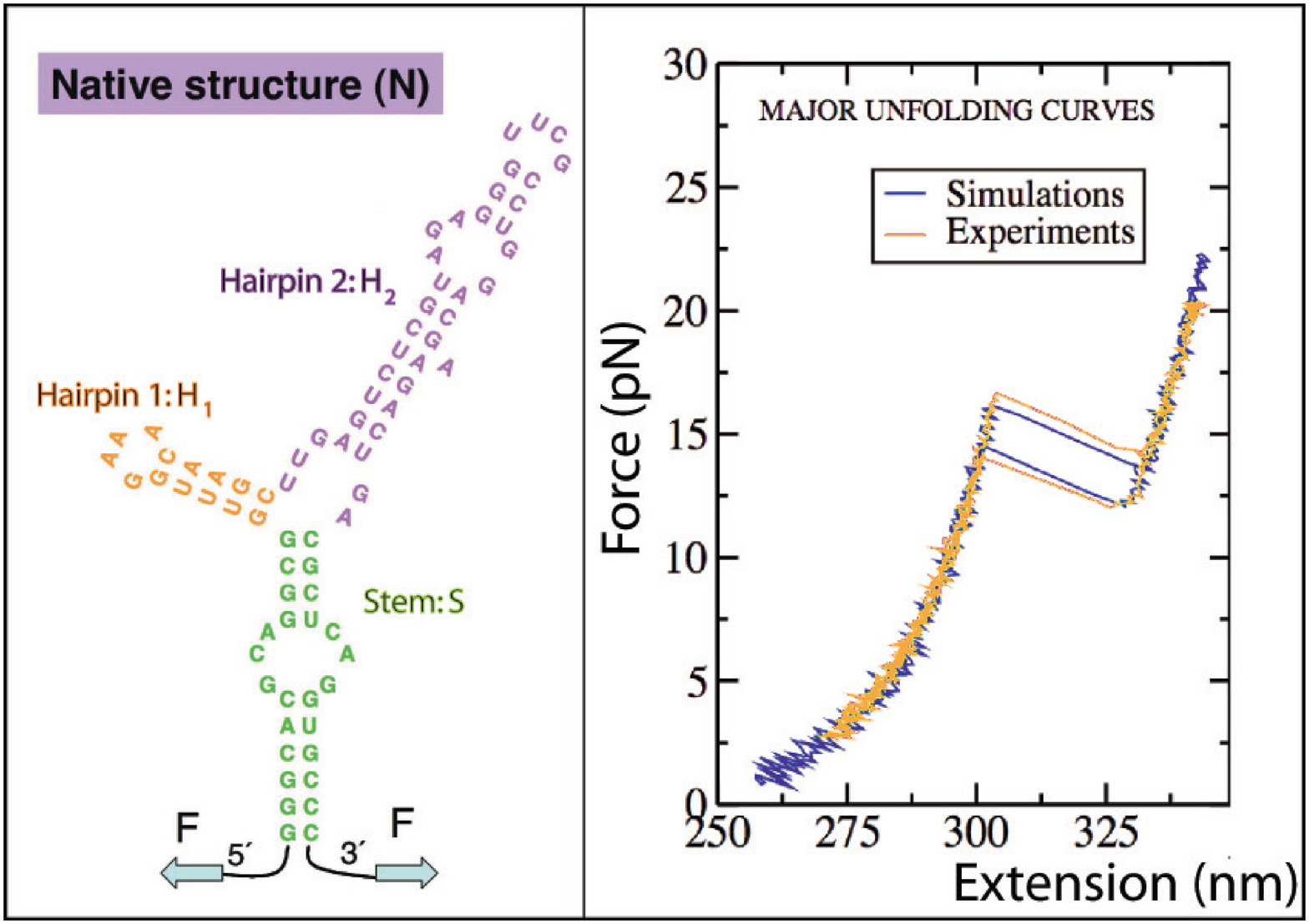}} 
\caption{\label{f2}} 
\end{figure}

As shown in Fig. \ref{f2}, the major unfolding curves are well  
reproduced by using an extension of the sequential kinetic model  
introduced by Cocco {\it et. al} \cite{Cocco,Cocco2}, applied to the native  
three-helix junction (denoted by $N$). The model  
in \cite{Cocco} describes the folding/unfolding force kinetics of  
single hairpins at the level of individual base pairs. It has one free  
parameter which is the attempt frequency, $k_a$, for the opening and  
closing rates of a single base pair (see Methods).  We extend this  
model to include multi-branched structures such as  
$N$ in RNAS15, which is composed of a stem $S$ that  
branches into two hairpins $H1$ and $H2$ (Fig. \ref{f2}). We  
also include the effect of the instrumental setup used in the  
optical tweezers experiments \cite{man4}.

\begin{figure}[t] 
\centerline{\includegraphics[scale=0.5]{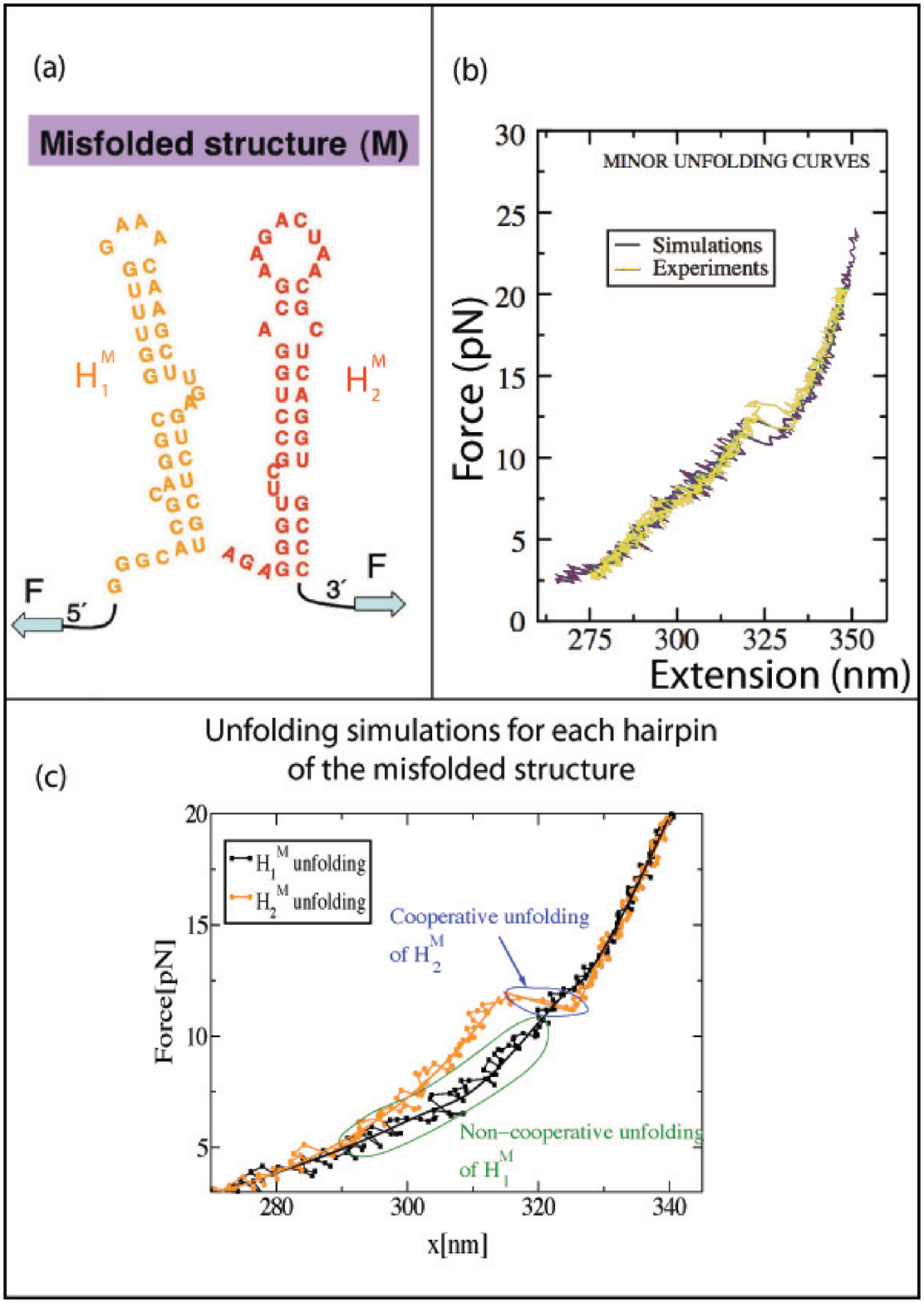}} 
\caption{\label{f3}} 
\end{figure}

Our numerical results show that, during the unfolding transition, the  
whole structure unfolds immediately after the stem opens. Accordingly,  
an analysis of the distribution of breakage forces predicts a  
transition state for the unfolding reaction that is located close to  
the native state (see Methods). The corresponding kinetic barrier is  
actually generated by the presence of successive strong GC  
base pairs in the stem. On the other hand, this sequential dynamics  
applied to the structure composed of the native hairpins $H_1$ and  
$H_2$ does not reproduce the minor curves (data not shown). This  
suggests that the minor curves correspond to the unfolding of a  
misfolded structure, rather than to the unfolding of a structure that  
is partially folded into $N$ (with hairpins $H_1$ and $H_2$, but not  
the stem, formed). By using the Vienna package for predicting RNA  
structures \cite{Vienna} we have searched for the most stable  
structure without the stem formed (in order to avoid the large  
cooperative rip characteristic of the major curves).  This structure,  
denoted as $M$, is composed of two hairpins, $H_1^M$ and $H_2^M$, and  
has a free energy of 6.3 kcal/mol ($\simeq 10.5 k_B T$) above that of  
the native structure (see Methods and Fig. \ref{f3}a) --note that $N$  
and $M$ cannot coexist at the same time since the same nucleotides are  
involved in different base pairings. Upon stretching $M$, numerical  
simulations show minor-like unfolding curves similar to the  
experimental ones (see Fig. \ref{f3}b). In the simulations, the  
cooperative transition observed around $14$ pN corresponds to the  
unfolding of the $\sim 30$ bases hairpin $H_2^M$ as shown in  
Fig. \ref{f3}c. This figure also shows that for loading rates similar  
to those of the experiments, $H_1^M$ unfolds in a non-cooperative way  
at force values between 6 and 9 pN (see Appendix \ref{app1} for a discussion on this issue). This corresponds to the non-cooperative  
transition observed in the experimental (unfolding) minor curves  
(see above and Fig. \ref{f1}). In the following, we provide  
quantitative evidence showing that the minor curves indeed result from  
the formation of $M$.  

\section{The minimal structures model}

In order to investigate the folding/misfolding in RNAS15 we introduce
a model that can be applied to any nucleic acid secondary
structures. We call it the {\bf\em minimal structures model} (MSM).
The essential idea behind the model consists in associating to each
type of experimental unfolding curve --two in the case of RNAS15,
``major'' and ``minor''-- a unique stable structure, whose unfolding
force-extension pattern, obtained using the sequential dynamics,
reproduces the experimental one.  From this set of stable structures,
that we call {\it minimal} structures, we generate the ensemble of
configurations used to investigate both the unfolding and the
refolding of the molecule.  These configurations, hereafter referred
to as MSM configurations, are built as follows.  First, we consider
all the intermediate configurations resulting from the sequential
unfolding of each minimal structure.  Each of these intermediate
configurations is composed of hairpins that are separated by regions
of unpaired bases. The ensemble of MSM configurations results from all
the possible combinations of these hairpins (Fig. \ref{f4}). The
initial set of locally stable structures is said to be {\it minimal}
since each of these structures is necessary to reproduce one of the
pattern of unfolding force-extension curves obtained in the
experiments. Moreover, this minimal set of structures makes
simulations of kinetics affordable form a computational point of view
(the number of configurations in the MSM grows in a polynomial way as
$\prod_{i=1,\#MS} N_{i}$, $N_{i}$ being the total number of base pairs
of the minimal structre $i$ and $\# MS$ the total number of minimal
structures). Although the inclusion of more structures might appear
desiderable, the implementation of the kinetics soon becomes
exceedingly complicated and little is actually gained regarding
comparison with the experiments.  Finally, the dynamics that we
implement at the level of single base pairs \cite{Cocco} satisfies
detailed balance and is ergodic (i.e. each configuration in the MSM is
connected through a path, made out of a finite number of successive
openings and closings of base pairs, to any other
configuration). Detailed balance and ergodicity are essential
properties of the dynamics ensuring that, in the equilibrium state,
all configurations are accessible and sampled according to the
Boltzmann-Gibbs distribution. Detailed balance and ergodicity make the
link between dynamics and thermodynamics where time averages can be
replaced by ensemble averages.

During refolding there is always competition in the formation of 
hairpins that have bases in common (e.g $H_1$ and $H_2^M$ in 
RNAS15). Therefore, with more than one minimal structure, the MSM 
naturally leads to the formation of the different minimal structures and 
hence to misfolding. In RNAS15, comparison between experiments and 
numerical simulations for the unfolding curves (Fig.  \ref{f2} and \ref{f3}) 
suggests to choose $N$ and $M$ as the minimal structures. The total 
number of configurations within the MSM being on the order of a few 
hundreds. We have 
carried out numerical simulations of force cycles in the MSM in RNAS15 and 
observed the presence of minor and major unfolding curves in agreement 
with the experiments. Yet, the current model is not good enough to 
reproduce the experimental results as we are still not able to 
simultaneously reproduce the unfolding and refolding curves in a 
quantitative way (data not shown). In particular, by choosing a value of 
the attempt frequency $k_a$ that fits well the unfolding curves, we 
obtain refolding curves that do not match the experimental results 
(typical refolding forces are 2 pN higher in simulations than in 
experiments). Different causes could explain this discrepancy. First, we 
have neglected a large number of configurations that might compete with 
those of the MSM and whose presence would lead to lower refolding forces 
in agreement with the experimental results. In addition, the transient 
formation of tertiary interactions such as pseudo-knots, could be 
relevant during the folding process. 

The number of secondary structures that can be formed in RNA grows
exponentially with the total number of bases. Therefore, it is
impossible, in large molecules, to simulate kinetics in the full
ensemble of secondary structures. Although it is possible to determine
the free energy of all possible secondary structures it appears
extremely difficult to implement kinetic rules between all possible
configurations.  The simplest strategy, in order to include the effect
of additional structures on the dynamics, is to consider all possible
secondary contacts that can be formed within the unpaired regions in a
given MSM configuration. Because the explicit inclusion
 of all possible secondary structures in the dynamics is too difficult,
 we take advantage of approximative schemes to address such
 problem. The current problem is reminiscent of that encountered in
 liquid or statistical field theories where an infinite class of
 correlation functions or observables have to be simultaneously
 solved. It is then common to solve the dynamics by closing the
 hierarchies of observables by selecting only a specific subset among
 all possible classes and resumming all diagrams among that
 subset. Here we adopt such strategy. In the spirit of resummation
 techniques in statistical physics, we integrate out all these additional
 structures and add corrections to the free energies of the MSM
 configurations as explained below.

\subsection{ Estimate of the free-energy correction in the MSM.}  

\begin{figure}[t]  
\centerline{{\includegraphics[scale=0.5]{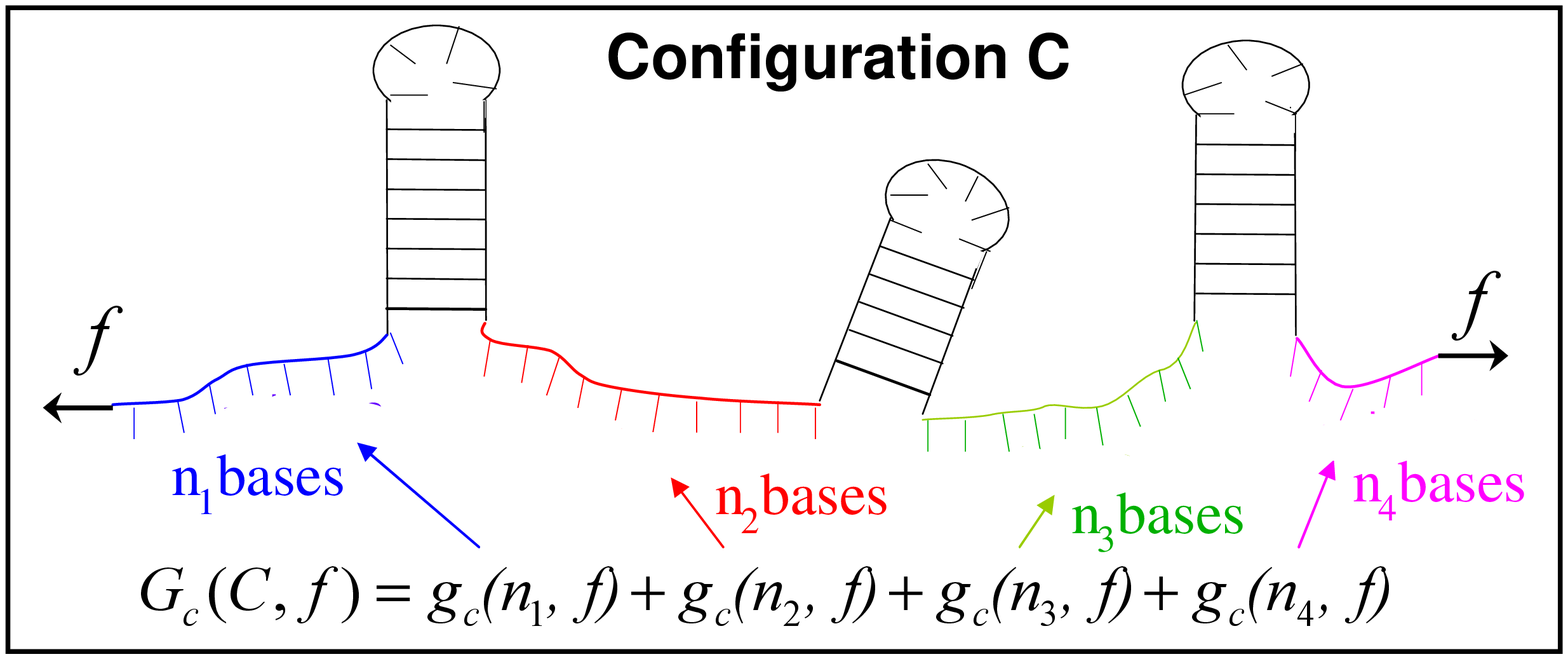}}}  
\centerline{{\includegraphics[scale=0.5]{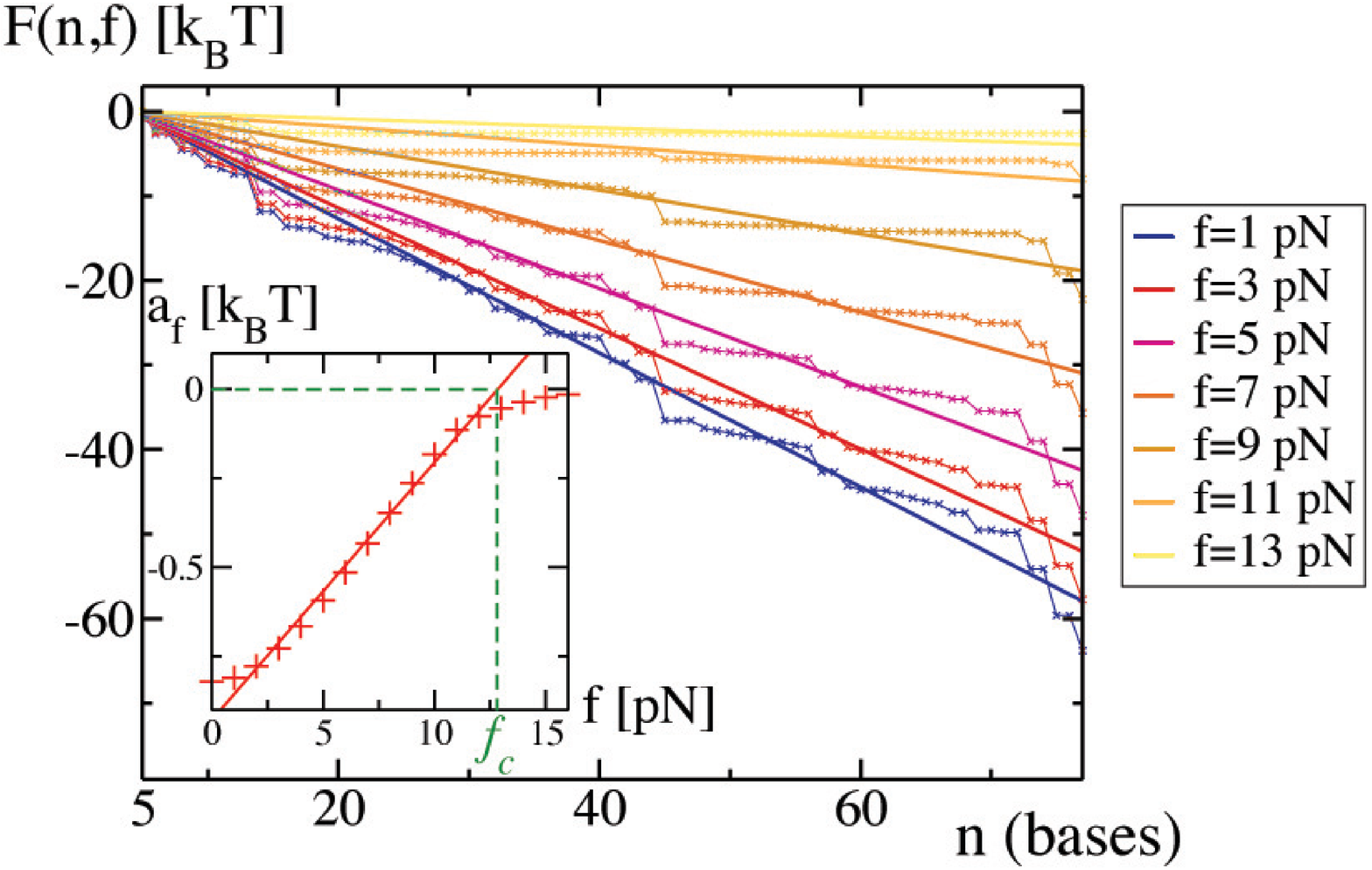}}}  
\caption{\label{fbox2}}
\end{figure}

Let us consider a generic configuration $C$ of the MSM with free
energy $G(C,f)$ at a given force $f$. $C$ is by definition composed of
hairpins and regions of unpaired bases (Fig. \ref{fbox2}).  Starting
from this configuration, we can generate additional ones by allowing
the formation of secondary contacts between complementary bases within
each unpaired region. The inclusion of these additional configurations
in the MSM would result in a larger ensemble of configurations. This
would also modify the thermodynamics of the system. Hence, in order to
keep an ensemble of configurations as small as possible, the effect of
such additional configurations is taken into account by adding a free
energy correction, $G_c(C,f)$ to each configuration
$C$. Subsequently, the free energy of any configuration $C$ in the
 MSM can be split into three contributions:
\begin{equation}  
G(C,f)=G_{0}(C)+G_m(C,f)+G_c (C,f).  
\end{equation}  
$G_{0}(C)$ is the free energy of formation of the configuration  
 $C$ at zero force. $G_m (C,f)$ stands for the contribution to the 
 mechanical free energy due to the stretching of the unpaired regions 
 that are exposed to the force. This is equal to $\int_0^f 
 x_{C}(f')df'$ where $x_{C}(f)$ is the equilibrium average extension of 
 the configuration $C$ at force $f$. Finally, the free energy 
 correction at force $f$, $G_c(C,f)$, is added so that $G(C,f)$ includes  $C$ and all the possible 
 secondary structures that can be formed from $C$ using the bases of the 
 unpaired regions. Note that some of these structures may correspond 
 to configurations originally belonging to the MSM and, therefore, should not be included in the calculation of 
 $G_c(C,f)$. In fact, the inclusion of such structures would lead to an 
 incorrect and strongly biased estimation of the free energy correction 
 inherent to the large thermodynamic stability of all configurations that belong to the MSM. The proper 
 estimation of $G_c(C,f)$ is therefore a very difficult task and a 
 different strategy is required to circumvent this problem as we shall explain in the following.

In the present treatment, for the sake of simplicity, we do not consider
interactions between bases of different unpaired regions. As a
consequence, $G_c(C,f)$ can be decomposed as a sum of independent
contributions $g_c^i$ coming from each unpaired region $i$. Having
proceeded so far, we try to get an estimation of the correction
$G_c(C,f)$ that can be efficiently implemented in the numerical
simulations of the kinetics. We use an annealed approximation where the
contribution from each region $i$ only depends on the number $n_i$ of
bases of that region, $g_c^i=g_c(n_i,f)$. As a result, we get
$G_c(C,f)=\sum_{i=1}^{N_U}g_c(n_{i},f)$ where $N_U$ is the total number
of unpaired regions (see Fig. \ref{fbox2}).

As the free energy of an RNA sequence depends much on its sequence, $g_c(n,f)$ should be estimated for each primary sequence. In this regard, our estimation procedure consists, first, in evaluating the average free energy of an $n$-base long polynucleotide chain that is chosen within that
sequence (see Methods). The average is taken over all possible segments
of length $n$ along that sequence.  To this value we subtract the
initial stretching free energy $G_m(n,f)$ of the $n$-bases long
polynucleotide and obtain $F(n,f)$.  $F(n,f)$ is always a lower bound to
$g_c(n,f)$ as it includes the contribution coming from the additional
new configurations but also the contribution from configurations already
generated by the minimal structures. In fact, by averaging over all
segments covering the whole sequence, the term $F(n,f)$ gets
contributions from all possible hairpins that can be formed with $n$
bases. Therefore $F(n,f)$ is biased toward low values due to the 
stabilizing contribution to the free energy by the minimal structures 
(e.g. the native or the misfolded structures in the case of RNAS15). This 
bias is particularly strong at low forces where the native hairpins 
dominate the annealed average.  How does $F$ depend on $n$ and $f$?  
The fact that the free energy $F$ is an extensive variable (i.e.
depends linearly on the size of the system $n$, at least for $n \geq
5$ where loop formation is possible) implies that the first derivative
$\partial F /\partial f$ (i.e. respect to the intensive variable $f$)
also depends linearly on $n$.  These properties are well confirmed by
using the Vienna package \cite{Vienna}, which gives the exact
partition function and the equilibrium free energy for any RNA
sequence. In the case of RNAS15 we find $F(n,f)\approx a_f (n-5)$
where the parameter $a_f$ depends linearly on $f$ up to a certain
force value $f_c \simeq 12$ pN for which it vanishes: $a_f\simeq
a(f-f_c)/f_c$ if $f<f_c$ and $a_f=0$ if $f \geq f_c$, with $a\simeq
0.5 {\rm kcal/mol}=0.9 k_BT$ (see figure \ref{fbox2}). We stress that,
for arbitrarily long sequences, determining $a$ and $f_c$ is still
possible by restricting the calculation of the free energy $F(n,f)$ to
small values of $n$ (e.g. up to $n\simeq 50$) where $a_f$ is a linear
function of $f$ (Fig.~\ref{fbox2}).

How to proceed now in order to estimate the true correction $g_c(n,f)$?
The functional form obtained for $F(n,f)$ suggests the same functional 
dependence for $g_c(n,f)$, albeit with {\it a priori} different 
parameters, $a$ and $f_c$. $f_c$ in $F(n,f)$ is the force 
value where the free energy correction vanishes and below which 
secondary structures become, in average, more stable than the fully 
unfolded or unpaired form.  At forces around $f_c\simeq 12$ pN many 
other configurations can be as stable as the MSM configurations. 
Therefore, the 
value of $f_c$ is not expected to be very sensitive to the bias 
introduced in the annealed average by the inclusion of the MSM 
configurations. Thus, we keep $f_c\simeq 12$ pN for $g_c(n,f)$ also. 
Consequently, the free energy correction term leads to only 
one additional free parameter in the model, that we call $A$ . The free energy 
correction finally reads $g_c(n,f)\approx A_f (n-5)$ with $A_f\simeq 
A(f-f_c)/f_c$ if $f<f_c$ and zero otherwise.  The parameter $A$ 
corresponds to the free energy correction per base pair at zero force 
and satisfies $A\le a$ because $F(n,f)$ is a lower bound to $g_c(n,f)$. 
What is the main 
effect of $A$ on the kinetics of unfolding and folding?. Additional 
configurations naturally tend to slow down the formation of individual 
hairpins that belong to the minimal structures.  Accordingly, the free 
energy correction modifies the closing rates rather than the opening 
rates of individual base pairs (see Methods). Therefore the value 
of the parameter $A$ mostly determines the kinetics of folding rather 
than unfolding and a larger value of $A$ tends to slow down the kinetics 
of folding. 

\subsection{Applying the model to RNAS15.}   
\begin{figure}[t]  
\centerline{\includegraphics[scale=0.5]{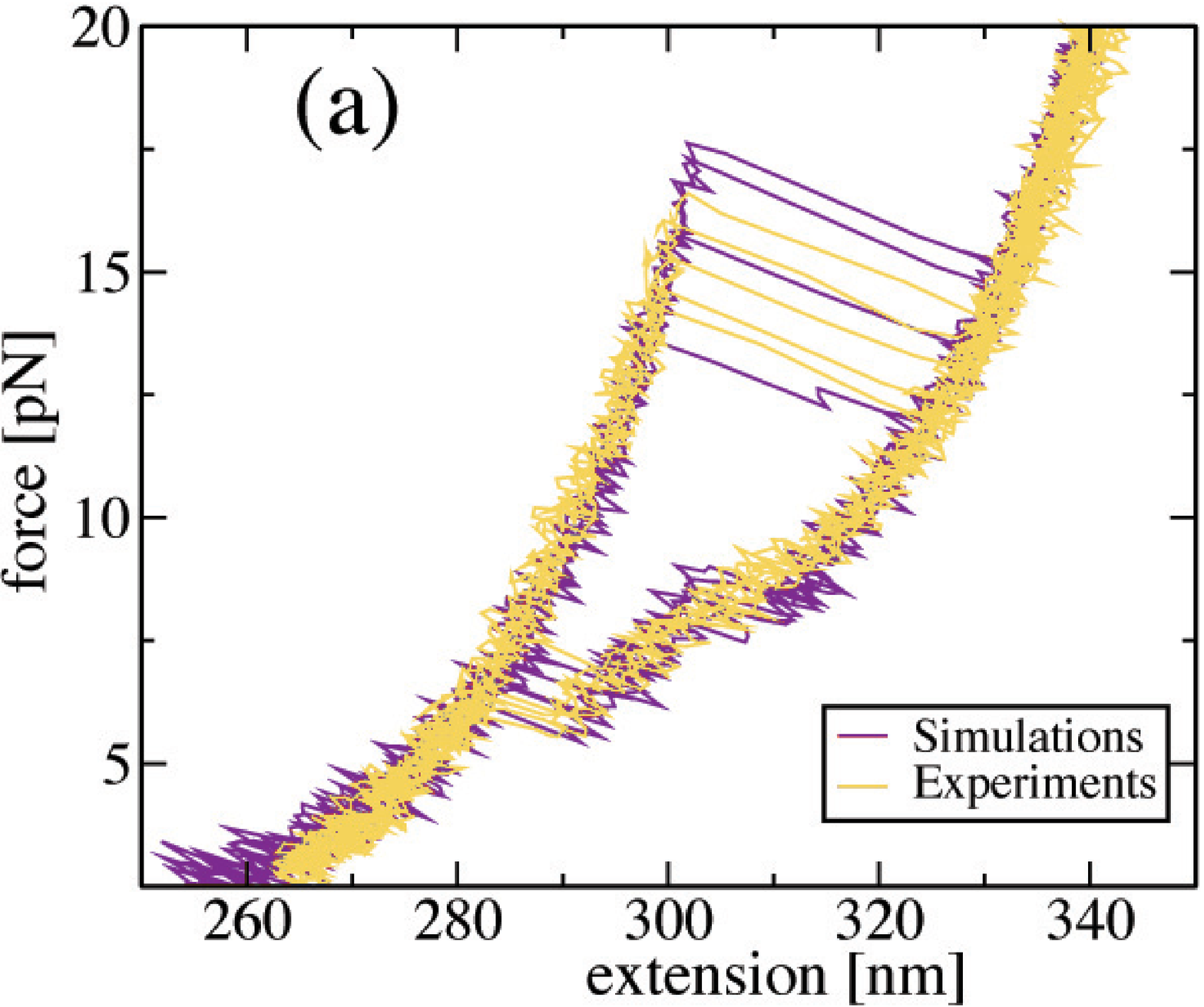}}  
\vspace{.5cm}  
\centerline{\includegraphics[scale=0.5]{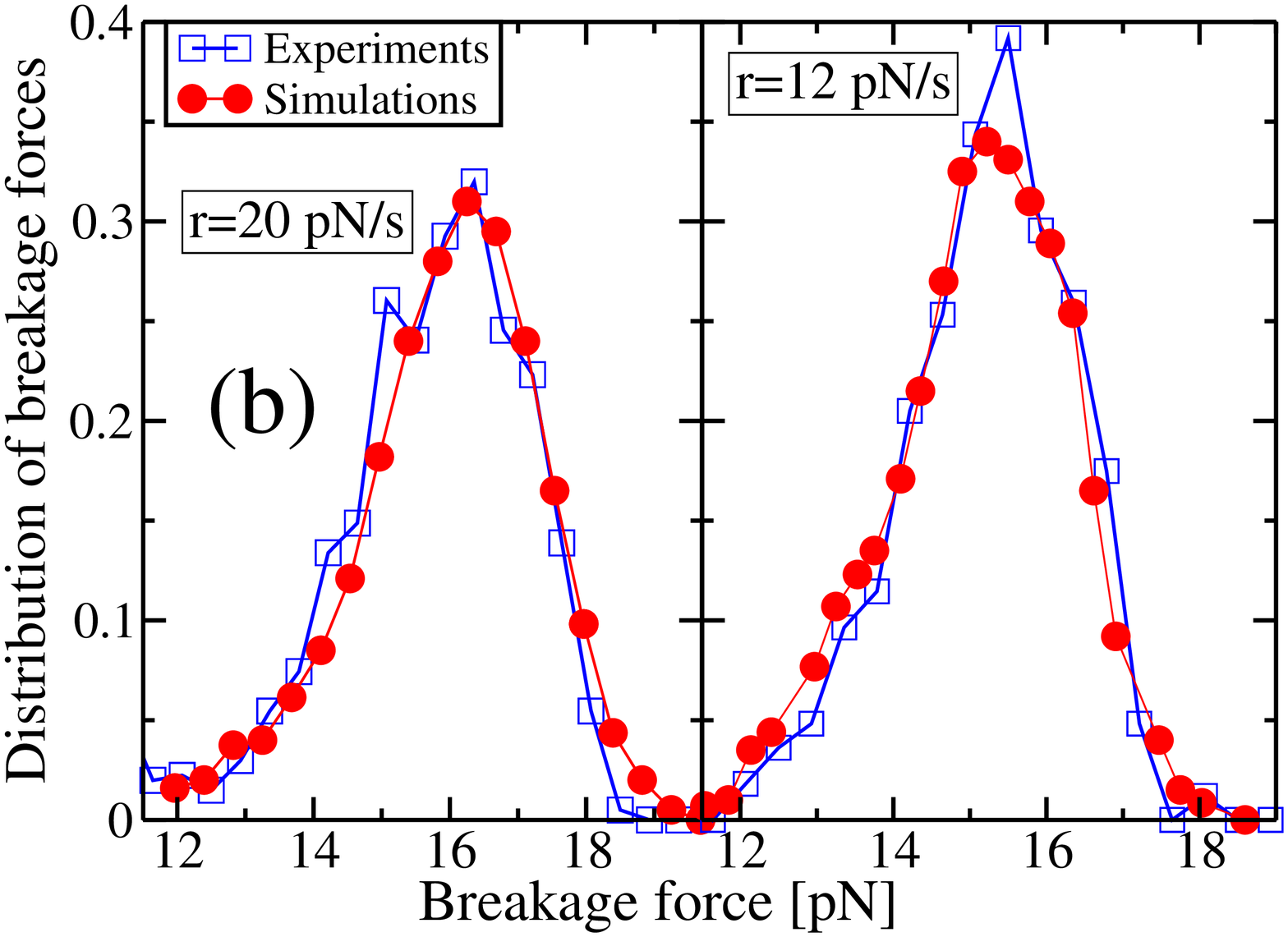}}  
\caption{\label{f5}}  
\end{figure}

Overall the model requires only two free parameters, $k_a$ and $A$, in 
order to fit all the experimental data available in RNAS15. The parameters 
$A=0.3 k_B T$ and $k_a=10^7 s^{-1}$ lead, at both loading rates, to 
unfolding and refolding force-extension curves, distributions of breakage force and 
misfolding probabilities that are in quantitative agreement with those 
found in the experiments (Fig. \ref{f5} and \ref{f6}).  Since no further 
explicit structures are necessary to reproduce the experimental data, we 
conclude that, in this case, a model containing the minimal structures 
$N$ and $M$ plus the free energy correction term, is enough to explain 
both the unfolding and refolding kinetics of RNAS15. In this regard, we 
have extended our analysis by including other minimal structures 
different from $N$ and $M$ and have obtained very similar results (data 
not shown). 

Regarding the force-extension curves we note that the shoulder observed during the  
refolding trajectory (Fig. \ref{f5}a) is mainly due to the transient  
formation of hairpins ($H_1$, $H_2$, $H_1^M$ and $H_2^M$). On the  
other hand, the minor curves correspond to the unfolding of the  
misfolded structure $M$ where the hairpin $H_2^M$ does not allow the  
formation of the native hairpin $H_1$: $M$ acts as a kinetic trap that  
impedes the formation of $N$. Misfolding in RNAS15 is not induced by  
thermal fluctuations since the free energy difference between $N$ and  
$M$ is very large, $\Delta \Delta G_0 \sim 10.5k_BT$. Rather it is  
induced by the force that tends to favor the misfolding pathway.

Finally, we note that the free energy correction per base pair, $A\simeq  
 0.3 k_B T$, is an order of magnitude smaller than the typical free  
 energy of formation of individual base pairs ($\sim 3 k_B T$). Yet,  
 it is necessary to include this correction (about 10\%) to 
 quantitatively reproduce 
 the experimental features of the unfolding/refolding kinetics in RNAS15.  

\section{Misfolding probability}

\begin{figure}[t]  
\centerline{\includegraphics[scale=0.5]{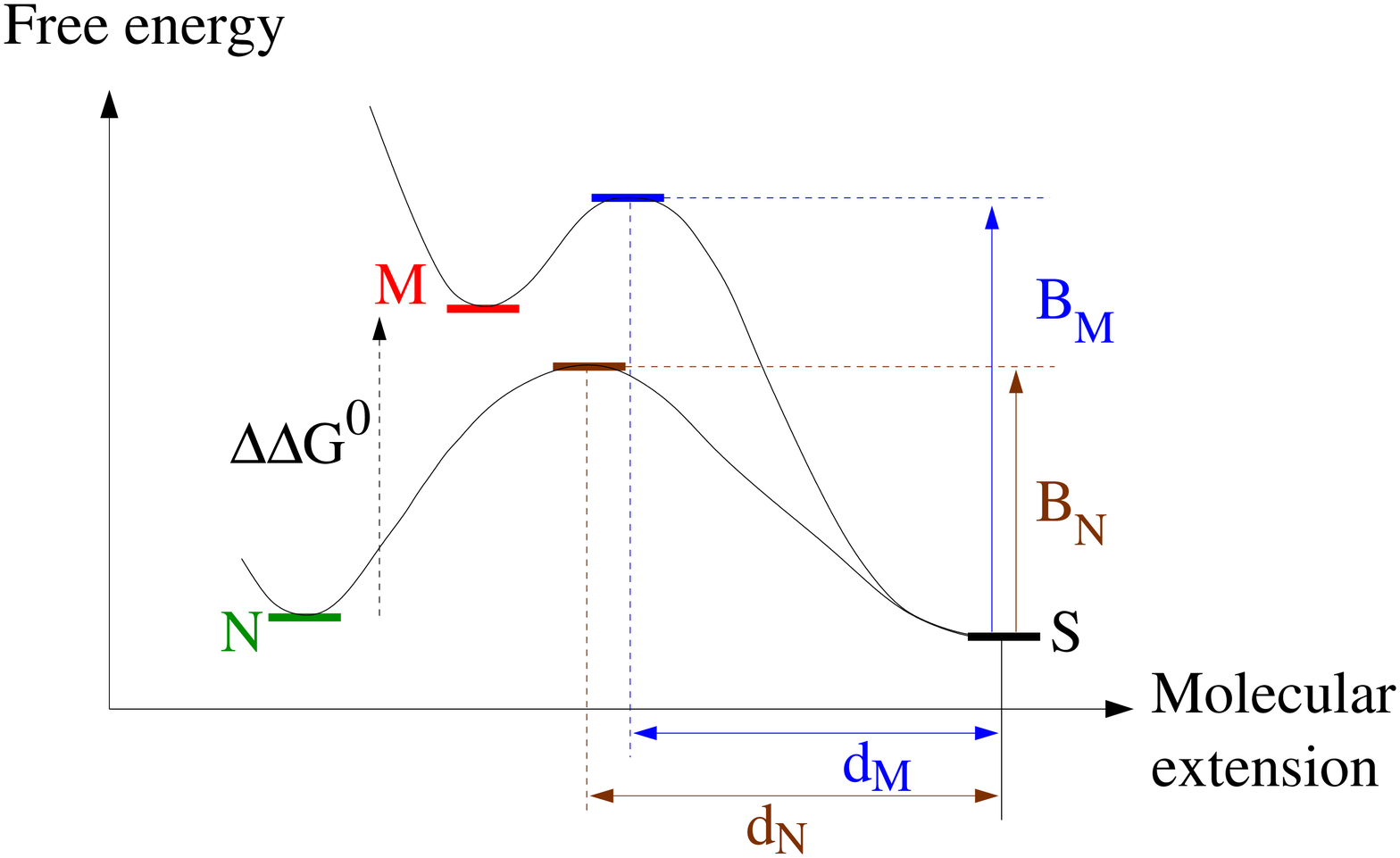}}  
\vspace{1.5cm}
\centerline{\includegraphics[scale=0.5]{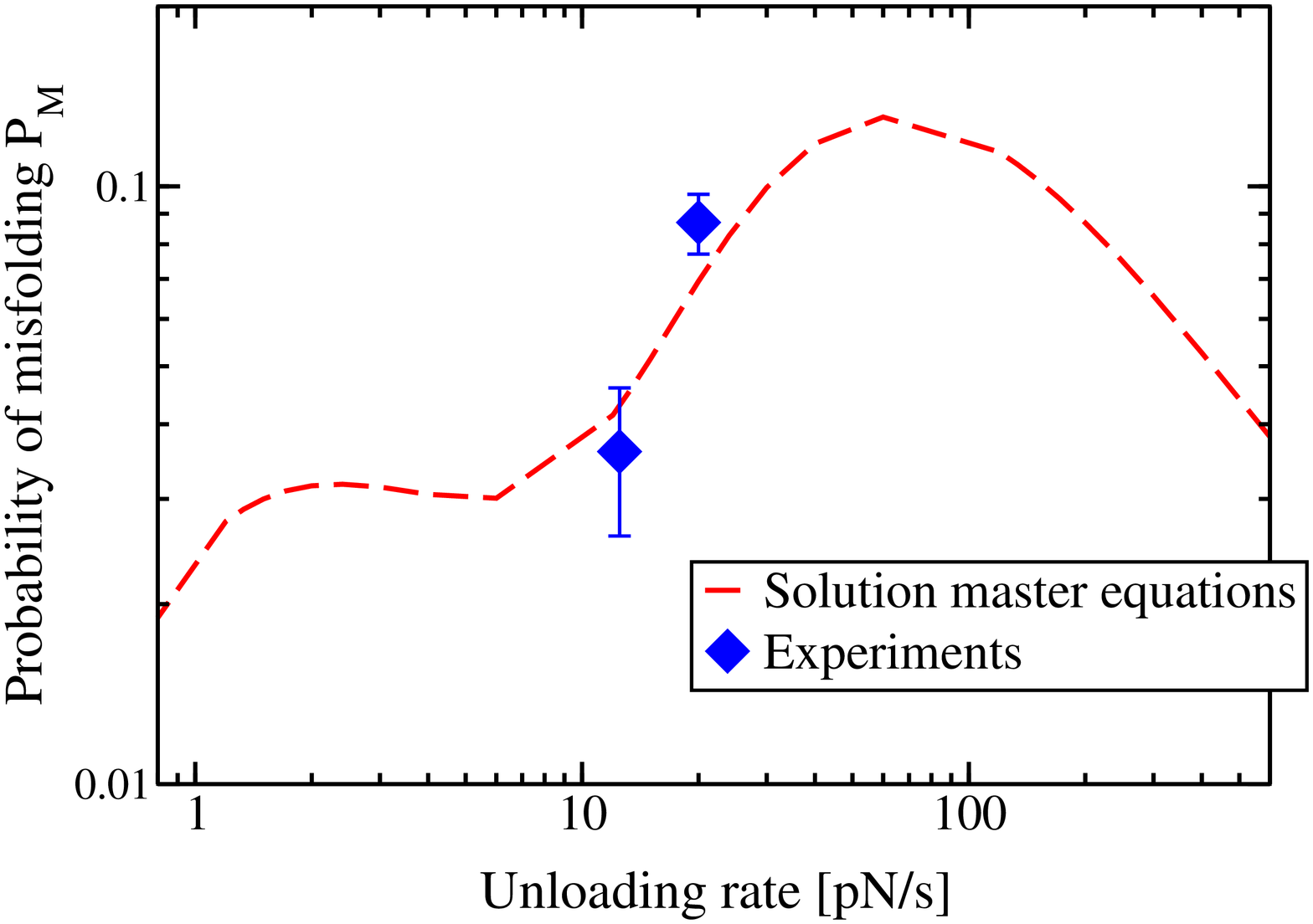}}  
\caption{\label{f6}} 
\end{figure}

In a force cycle protocol, misfolding can be quantified by the  
misfolding probability $P_M$. This is given by the probability to end up  
in the misfolded state at the end of the relaxing process. Multi-state  
models of chemical reactions provide a general picture about the  
unloading rate dependence of this probability. The simplest model  
consists of a three-state system (native ${\bf N}$, misfolded ${{\bf  
M}}$ and stretched ${\bf S}$) where the misfolded state ${\bf M}$ acts  
as a kinetic trap during the folding transition  
(Fig.~\ref{f6}a). Starting from ${\bf S}$ at high forces, and by  
decreasing the force at a constant rate $r$, the general question we ask  
is how $P_M(r)$ depends on $r$. In the general situation of a  
force-independent position of the kinetic barriers $B_N,B_M$ (located at  
distances $d_N,d_M$ from ${\bf S}$), we find that $P_M(r)$ has a unique  
maximum located at $r^*$ (see Appendix \ref{app2}). However, if $d_N$ and  
$d_M$ depend on the force, $P_M(r)$ shows a more complex behaviour  
where several maxima can appear (see Appendix \ref{app2}). This general  
scenario is expected to be applicable in RNAS15 where the results  
obtained from simulations of the MSM show a $P_M(r)$ with two maxima  
(Fig.~\ref{f6}b). From a general point of view, a $P_M(r)$ with more  
than one maximum suggests a complex free energy landscape with force  
dependent transition states (leading to force dependent fragilities as  
in the case of RNA hairpins \cite{man1}).

\section{Discussion and conclusions}

In this work we have investigated the folding/unfolding behaviour 
of nucleic acid secondary structures that are pulled by mechanical 
forces. To this aim we have introduced a phenomenological model (MSM) 
that is based on: the sequential dynamics of a minimal number of 
structures; and the inclusion of corrections in the free energy 
that account for the configurational entropy contributed by the 
exponentially large number of neglected secondary structures. The model 
describes force-induced misfolding of nucleic acid secondary structures 
such as RNA and DNA. It can be applied to arbitrary nucleic acid 
sequences that can form different secondary structure and can be used to 
predict the phenomenology observed in dynamic force spectroscopy 
measurements (breakage force distributions, force-extension curves and 
misfolding probability). The applicability of the approach has been 
shown in the case of the RNA three-helix junction S15. 

The model can be also used in the prediction of different folding
kinetics scenarios by implementing different sets of minimal
structures. Sometimes the full applicability of the model may require
the previous experimental identification of the minimal set of
structures that generate the different patterns of force-extension
curves. Although the model cannot predict misfolding for a given
sequence it can be applied to identify possible misfolded states as
well as kinetic intermediates by doing systematic {\em in silico}
experiments. A useful strategy could be using the Vienna package
\cite{Vienna} to build up the minimal set of structures and
consequently determine potential misfolded states by generating
different sets of secondary structures for the given RNA sequence.
Subsequently one should search for the most stable structures that can
be formed when native domains are not allowed. However, we are not
able yet to provide a receipt that leads to the systematic
determination of these states. As a consequence, the method we used
for the determination of the misfolded structure must be specifically
adapted to every RNA sequence.

For a given nucleic acid sequence the model only has two fitting
parameters, $k_a$ and $A$. The first one, $k_a$, is an attempt
frequency at the level of individual base pairs which should not vary
much with the specific sequence under study. In this regard, the value
we report for $k_a$ in RNAS15 is in agreement with the values obtained
for other RNA molecules \cite{Cocco,man4} as expected. The second
parameter, $A$, is a thermodynamic parameter related to the
configurational space of the molecule, i.e. the space of secondary
structures associated with a given nucleic acid sequence. In
principle, for a given RNA, the larger the ensemble of MSM
configurations, the smaller the correction, and hence the value of
$A$. However, the total number of configurations included in the free
energy correction grows exponentially with the total number of base
pairs of the molecule, whereas the number of configurations in the MSM
grows as a power of that total number.  Consequently, the inclusion of
more minimal structures in the model should not change much the value of $A$.
In addition, $A$ is the
free energy correction per base pair and, therefore, it should not be
much sensitive to the specific molecular sequence. Therefore it is
reasonable to expect that the reported value of $A\simeq 0.3k_BT$ is
largely constant among all RNA sequences under identical environmental
conditions (e.g. temperature and salt).  What happens in the case of
short canonical (i.e. fully complementary or Watson-Crick base-paired)
hairpins?  These molecules show two-state behavior and cooperative
folding \cite{Cocco,man4}, yet the entropic correction might still be
necessary to fully describe the kinetics of folding.  In this case,
there will be just one minimal structure (the native one) so the effect
of the entropic correction, albeit small, could be experimentally
observable. It would be very interesting to carry out future
experiments capable of identifying, in generic two-state molecules, this
correction of entropic origin. Finally, let us mention that a different
theoretical approach is required to model the thermal
denaturation of RNAs and the associated folding and misfolding
mechanisms. In this case, the dissociation of base pairs is not a
sequential process anymore.

Recent pulling experiments in TAR RNA \cite{LiBusTin07} have shown how 
stretching forces can help the formation of the native structure when 
the molecule is initially trapped in misfolded structures.  Here, we 
have found that a mechanical force can also induce the opposite effect, 
by favouring misfolding pathways that are unlikely in the absence of 
force. It remains a challenge to apply this model to predict the 
detection of misfolded structures and kinetic intermediates in single 
molecule pulling experiments for specifically designed nucleic acid 
sequences.

\section{Methods}

\subsection*{Optical tweezers experimental setup.}  

Experiments in RNAS15 were reported in a previous paper by Collin {\it  
et. al} \cite{Coll}.  Buffer conditions were 100 mM Tris-HCl, pH 8.1,  
1 mM EDTA, free of magnesium ions, at room temperature $T=298 K$.  
RNAS15 is attached, via RNA/DNA handles ($\simeq 160$ nm), to two  
micron-sized polystyrene beads.  One bead is held fixed at the tip of  
a micropipette. The force is measured through the detection of the  
light deflected by the bead in the optical trap (Fig. \ref{f1}).  

\subsection*{Transition state along the unfolding pathway.}  

From the breakage force data, one can obtain information about the  
transition state corresponding to the force-induced unfolding pathway using a  
two-state model.  According to this model, the variance  
$\sigma_{f}$ of the breakage force distribution is inversely  
proportional to the distance $x^{F}$ from the transition state to the  
folded native state, that is $\sigma_{f}=\frac{k_{B}T}{x^{F}}$.    
In RNAS15, this relation leads to a transition state for  
the unfolding reaction that corresponds to a configuration where only the  
first two or three base pairs of the stem are opened.  

\subsection*{Extended sequential dynamics.}  

\begin{figure}[t]  
\centerline{\includegraphics[scale=0.7]{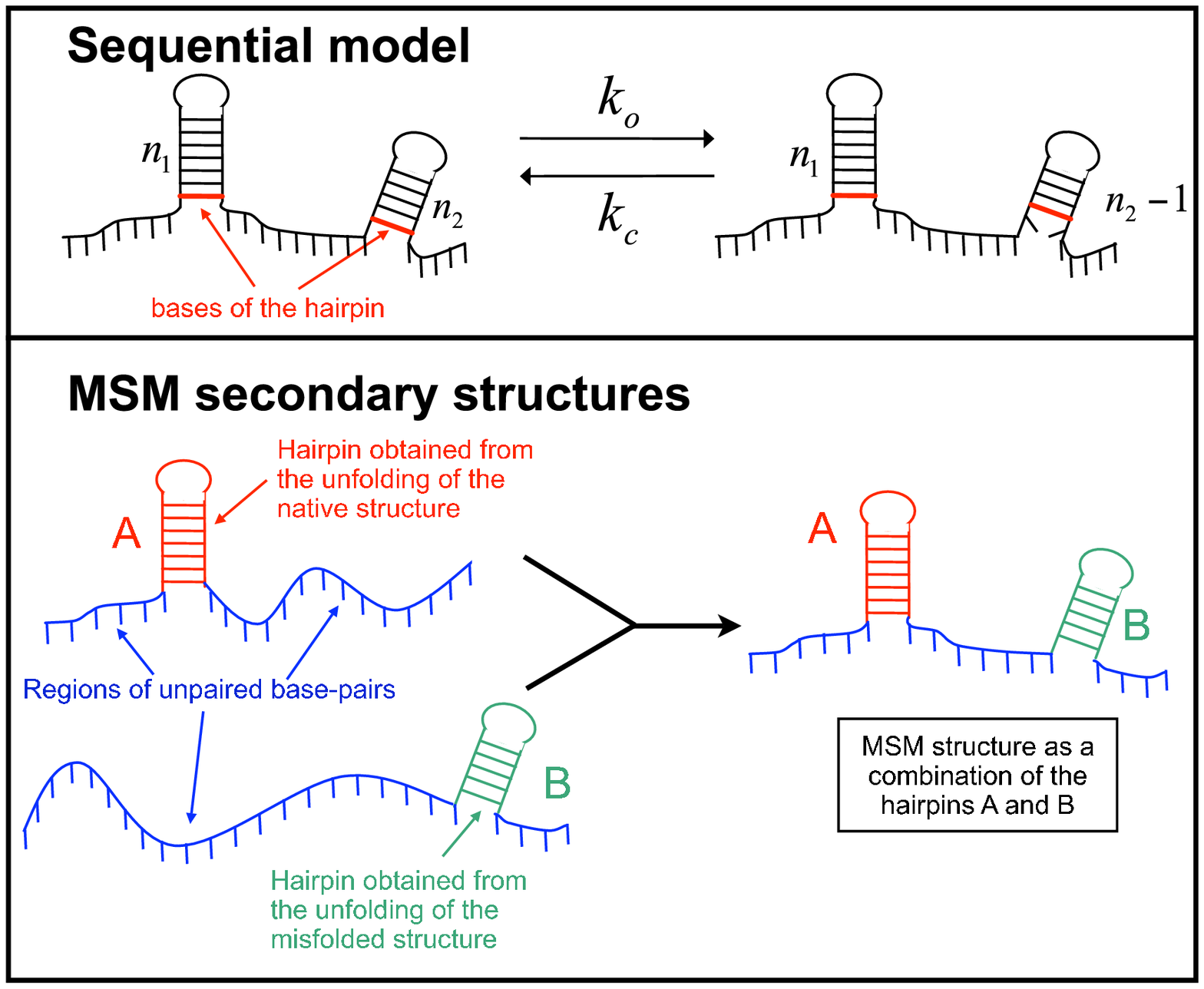}}  
\caption{\label{f4}}
\end{figure}

In the sequential model of Cocco {\it et. al} \cite{Cocco}, successive  
closing and opening of base pairs is restricted to take place at the  
base of the hairpin, defined as the first 5'-3' base pair formed  
(Fig. \ref{f4}). The corresponding opening rates ($k_o$) depend on  
the free energy of formation of the base pairs, $\Delta G_0$: $k_o=k_a  
\exp(- \Delta G_0/k_BT)$ where $k_a$ is an attempt frequency. The  
closing rates ($k_c$) depend on the mechanical energy loss, $  
\Delta G_m$, due to the shortening of the unpaired part of the  
molecule: $k_c=k_a \exp(-\Delta G_m/k_BT)$. These free energies have  
been estimated by thermal denaturation experiments \cite{Tin2} and  
single molecule force experiments respectively \cite{baumann,maier}. The  
attempt frequency $k_a$ is therefore the only free parameter of the  
model. Typical values measured by NMR fall in the range $10^7-10^8$  
Hz \cite{Leron}.  The extension of the model to multiple hairpins is  
depicted in Fig. \ref{f4}.

In our simulations, we allow for the formation of both Watson-Crick and  
non-canonical (GA and GU) base pairs. The values for the free energies  
of formation of the different base pairs have been obtained from the  
Vienna package (corresponding to 1 M NaCl \cite{Vienna}) by adding a  
uniform correction in order to meet the salt condition of the buffer  
used in the experiments (100 mM Tris-HCl). The salt correction is  
determined by imposing the value for free energy of formation in RNAS15  
to be equal to that recovered in the experiments \cite{Coll}.  The  
algorithm involves the whole experimental setup (handles and beads)  
within the so-called mixed ensemble where the control parameter is the  
distance between the optical trap and the immobilized bead \cite{man4}  
(rather than the force). Therefore, we include in $\Delta G_m$ the  
contribution of both the handles and unpaired RNA. The latter and  
the regions of unpaired RNA bases are described by using a worm-like  
chain model \cite{Flor,smith} with persistence lengths of $10$ nm  
(handles) and $1$ nm (RNA) and contour lengths of $0.26$ nm/bp (handles)  
and $0.59$ nm/base (RNA). These values fit reasonably well the  
experimental force-extension curves in the region where the handles are strecthed. Each  
hairpin contributes to the total extension with an additional extension  
of $\simeq 2$ nm. Finally, when taking into account our  
phenomenological corrections, $k_c$ becomes $k_c=k_a  
\exp(-(\Delta G_m+\Delta G_c)/k_BT)$ where $\Delta G_c$ is the  
difference in the free energy corrections between the open and  
closed configurations.

\subsection*{Free energy of an $n$-bases long segment of RNAS15.}  

Any secondary structure that is built up from an $n$-bases long  
polynucleotide can be seen as a succession of unpaired regions and  
partial secondary structures closed by a base-pair (for instance, in  
Fig. \ref{fbox2} the partial secondary structures are the  
hairpins). The free energy of such secondary structure can then be  
divided into the mechanical free energy corresponding to the  
stretching of both the unpaired regions and the base-pairs that close  
the partial secondary structures, plus the free energy formation of  
each partial secondary structure. In RNAS15, we estimate the latter using  
the Vienna package. Computing the free energy of all the secondary  
structures that can be formed with the $n$-bases long polynucleotide  
allows us to determine the partition function, and hence the free  
energy, of the $n$-bases long polynucleotide at force $f$.

\subsection*{Misfolding probability in RNAS15.}  

We describe the dynamics of the MSM using a set of master equations  
(see see Appendix \ref{app3}).  These equations describe the time evolution of the  
probability of the RNA to be in a specific MSM configuration.  To get  
the misfolding probability we numerically integrate the set of  
equations.  The force is decreased at a given unloading rate $r$,  
starting from the stretched state at an initial force $f_{in}=20$  
pN. The misfolding probability is computed at the end of the  
relaxing process when the force vanishes, i.e. when $t=20/r$.

\paragraph{Acknowledgement.} We thank D. Collin and I. Tinoco Jr. for discussions during the initial stages of this work and M. Palassini and P. T. X. Li for  
useful comments on the manuscript.  I. J acknowledges financial  
support form the European network STIPCO, Grant No.  
HPRNCT200200319. F. R acknowledges financial support from the Spanish  
Research Council (Grants FIS2004-3454, NAN2004-09348) and the Catalan  
Government (SGR05-00688).

\paragraph{Corresponding author.} Requests for material   
should be addressed to F. R (ritort@ffn.ub.es)

\appendix

\section{Appendix: Cooperative unfolding of hairpins $H_{1}^{M}$ and  $H_{2}^{M}$ \label{app1}}

\begin{figure}[t]
\centerline{\includegraphics[scale=0.5]{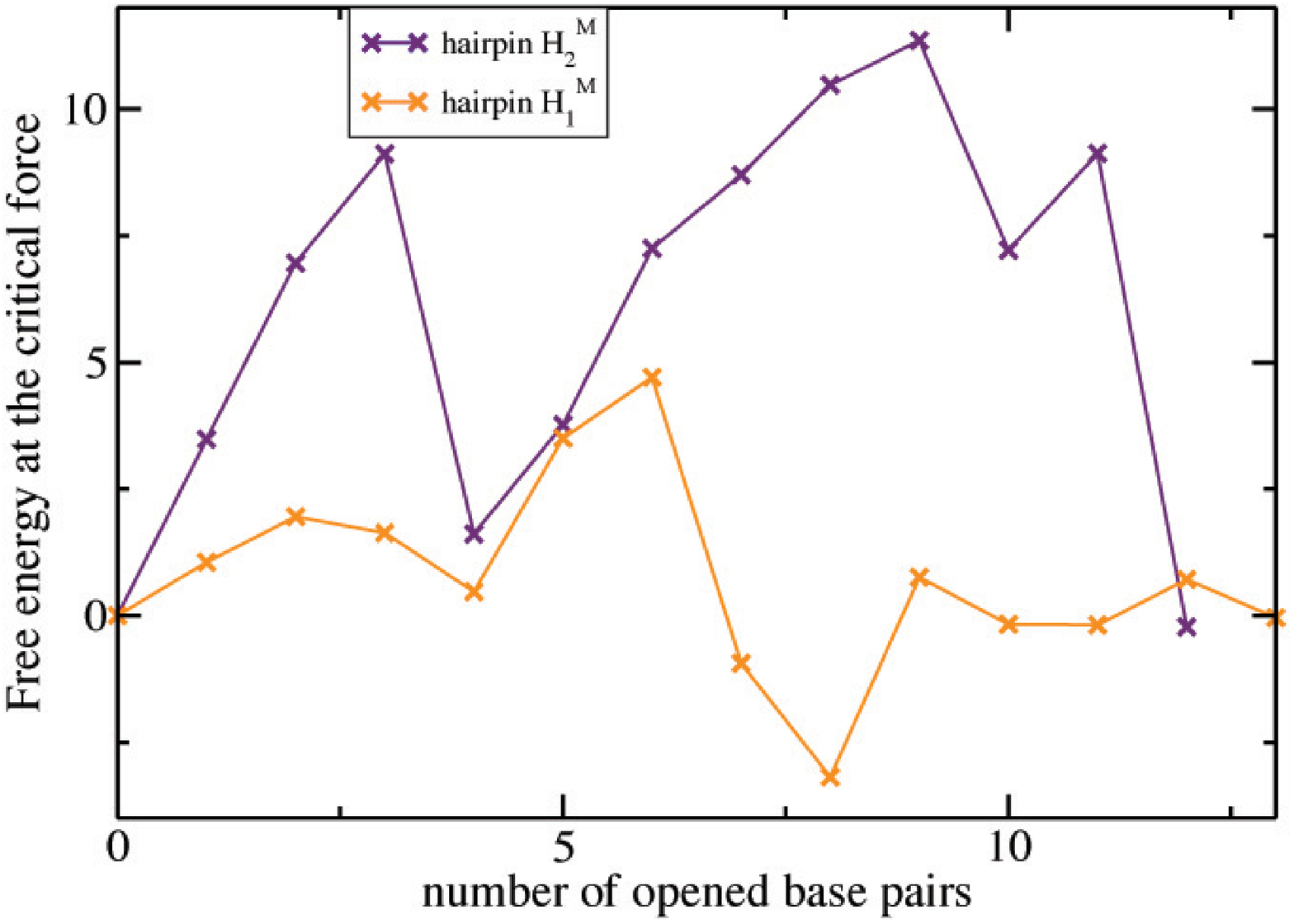}}
\caption{\label{ffree}}  
\end{figure}

The misfolded structure $M$ is composed of two hairpins $H_{1}^{M}$ and
$H_{2}^{M}$.  Both hairpins have similar thermodynamic stabilities and
they present several mismatches (internal loops and bulges).  Why
$H_{2}^{M}$ unfolds cooperatively whereas $H_{1}^{M}$ does not (see
Fig. \ref{f3}c)?  By using the Vienna package \cite{Vienna} for
the free energies of formation of different base pairs we can compute
the free energy of $H_{1}^{M}$ and $H_{2}^{M}$ as a function of the
number of denaturated base pairs at the critical force where the folded
and the unfolded hairpin are equally stable (i.e where both states have
the same free eenergy).  As shown in Fig. \ref{ffree} the free energy
landscape associated to $H_{2}^{M}$ (blue) presents a high kinetic
barrier between the folded and the unfolded hairpin, whereas the free
energy landscape associated to $H_{1}^{M}$ (red) is roughly flat.  This
explains the difference in the cooperativity observed between the two
hairpins.

\section{Appendix: Misfolding in a three-state model \label{app2}}

In this section, we analyse in detail the dynamics of a three-state 
model where a misfolded state (${\bf M}$) acts as a kinetic trap 
during the folding transition from the stretched state (${\bf S}$) to 
the native state (${\bf N}$). Let us consider the case of a  
pulling protocol where the mechanical force applied to the system 
decreases at a constant loading rate $r$. Starting from a high force 
value where the stretched state is the most stable one, we prove that 
the misfolding probability $P_M(r)$ at the end of the force releasing process 
shows a single maximum along the $r$-axis.

\begin{figure}[t]
\centerline{\includegraphics[scale=0.35]{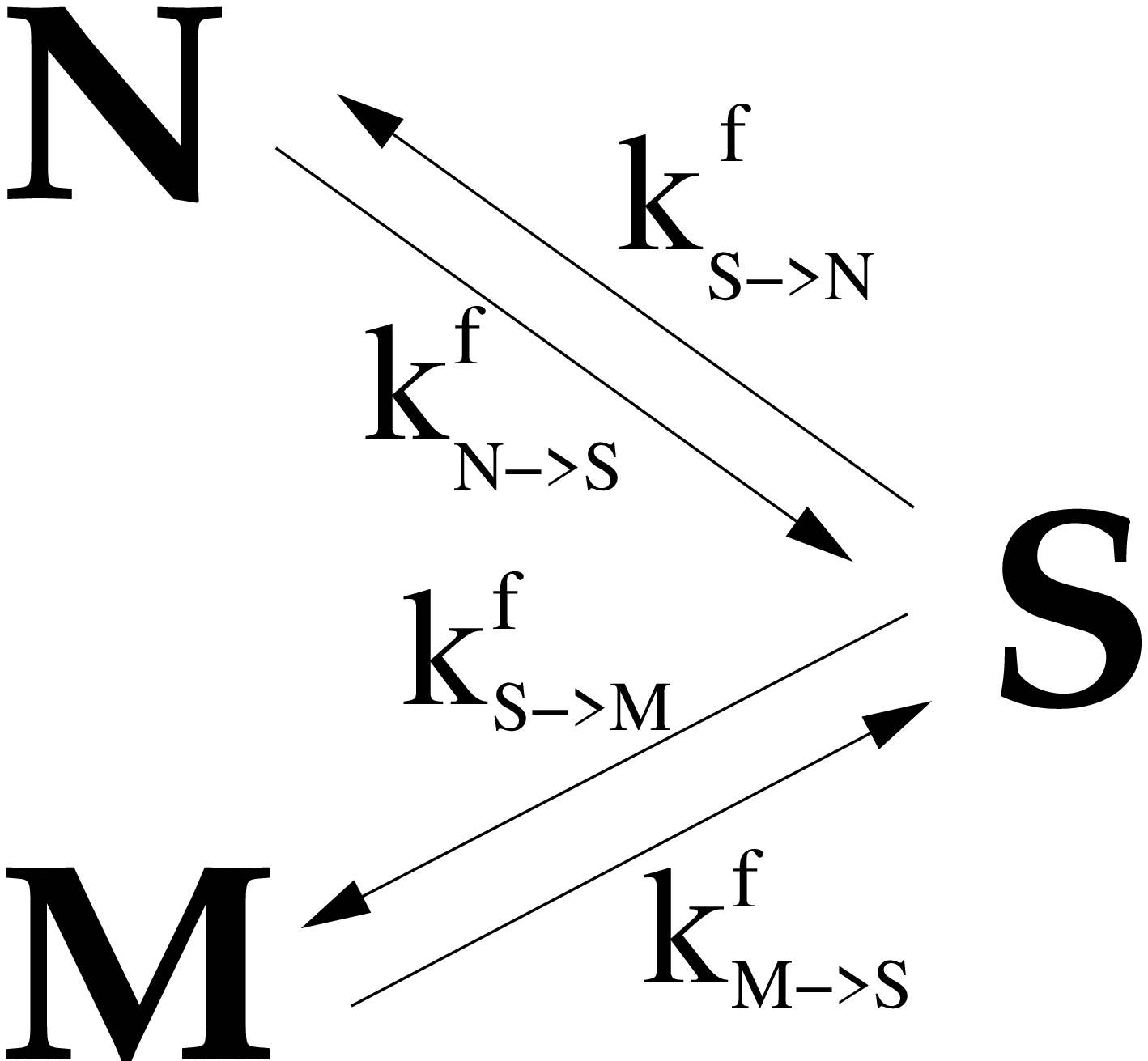}}
\caption{\label{MNS}}
\end{figure}

We denote by $P_N(t)$, $P_M(t)$ and $P_S(t)$ the probability to be at
time $t$ in the state ${\bf N}$, ${\bf M}$ and ${\bf S}$ respectively.
The relaxation process is governed by the following set of master
equations: 

\bea \nonumber \dot P_N&=&\frac{dP_N}{dt}=k_{S \to N}^fP_S-k_{N \to S}^fP_N\\ \nonumber
\dot P_M&=&\frac{dP_M}{dt}=k_{S \to M}^fP_S-k_{M \to S}^fP_M\\ 
\dot P_S&=&\frac{dP_S}{dt}=k_{N \to
S}^fP_N+k_{M \to S}^fP_M-(k_{S \to N}^f+k_{S \to M}^f)P_S \label{ME1}
\eea 

where $k_{a \to b}^f$ is the transition rate to go
from state $a$ to state $b$ at a given force $f$. Note that this model
does not allow for direct transition pathways connecting ${\bf N}$ and
${\bf M}$. Transitions between these states always pass through the
stretched state ${\bf S}$. ${\bf S}$ can then be viewed as an obligatory
intermediate state of the reaction ${\bf N} \rightleftharpoons {\bf M}$
(see Fig. \ref{MNS}).

\subsection*{Absorbing states}

In a first stage, we study the analytically tractable case where ${\bf 
N}$ and ${\bf M}$ are absorbing states, i.e. $k_{N \to S}=0$ and $k_{M 
\to S}=0$. The set (\ref{ME1}) of master equations becomes: 
\bea 
\nonumber 
\dot P_N&=&k_{S \to N}P_S\\ 
\nonumber 
\dot P_M&=&k_{S \to M}P_S\\ 
\dot P_S&=&-(k_{S \to N}+k_{S \to M})P_S 
\label{ME2} 
\eea 
In the presence of a mechanical force that is coupled to the molecular 
extension, the rates $k_{S \to N}, k_{S \to M}$ can be 
written as $k_{S \to N}=k_N \exp(-\beta d_N f)$ and $k_{S \to M}=k_M 
\exp(-\beta d_M f)$ respectively, where $d_N$ ($d_M$) is the distance  
along the reaction cooordinate between ${\bf S}$ and the kinetic barrier separating the state ${\bf 
 S}$ from the state ${\bf N}$ (${\bf M}$) (see Fig. \ref{f6}), $k_N$ and 
$k_M$ are the rates at zero force respectively and $\beta=(k_B T)^{-1}$ is the inverse of the thermal energy unit. Using 
these relations for the rates and considering a ramping protocol where 
the force decreases at a constant rate $r$ ($\dot f=-r$), the set of equations (\ref{ME2}) 
can be written in terms of the force as follows: 
\bea  
\nonumber  
\frac{dP_N}{df}&=&-\frac{k_N}{r}e^{-\beta d_N f} P_S\\  
\nonumber  
\frac{dP_M}{df}&=&-\frac{k_M}{r}e^{-\beta d_M f} P_S\\  
\frac{dP_S}{df}&=&\frac{1}{r}(k_Ne^{-\beta d_N f}+k_Me^{-\beta d_M f})P_S. 
\label{ME3} 
\eea 
Starting from an initial stretched state at very large force ($f\approx \infty$, $P_S=1$, $P_N=P_M=0$), the 
solution to (\ref{ME3}) is given by: 
\bea  
\nonumber  
P_S(f)&=&\exp\left(-\frac{k_Ne^{-\beta d_N f}}{r\beta 
 d_N}-\frac{k_Me^{-\beta d_M f}}{r\beta d_M}\right)\\ 
\nonumber  
P_N(f)&=&\frac{1}{r}\int_f^{\infty}dg\; 
k_N\exp\left(-\beta d_N g-\frac{k_Ne^{-\beta d_N g}}{r\beta 
 d_N}-\frac{k_Me^{-\beta d_M g}}{r\beta d_M}\right)\\  
P_M(f)&=&\frac{1}{r}\int_f^{\infty}dg\; 
k_M\exp\left(-\beta d_M g-\frac{k_Ne^{-\beta d_N g}}{r\beta 
 d_N}-\frac{k_Me^{-\beta d_M g}}{r\beta d_M}\right) 
\label{ME4} 
\eea 
Let us focus now on the misfolding probability 
$P_M=P_M(f=0)$.  
Starting from Eq. (\ref{ME4}) and after some simple 
manipulations, $P_M$ can be written as: 
\be 
P_M=P_M(\tilde r,\lambda,x)=\frac{1}{\tilde r}\int_0^{1} ds 
\;\exp\left(-\frac{s+\lambda s^x}{\tilde r}\right), 
\label{PM} 
\ee 

where $\lambda=\frac{k_N d_M}{k_M d_N}$, $\tilde r=\frac{r\beta 
d_M}{k_M}$ and $x=d_N/d_M$ are adimensional parameters. Interestingly, 
depending on the ratio $x=d_N/d_M$, two behaviors can be distinguished 
for the dependence of $P_M$ as a function of the adimensional rate $\tilde r$, 
i.e. of the rate $r$. In the following, we show that for $x <1$, $P_M$ 
has a single maximum along the $\tilde r$-axis, whereas for $x\geq1$, 
$P_M$ is a decreasing function of $\tilde r$. 

The first derivative of 
$P_M$ with respect to $\tilde r$ reads:  
\be 
\partial_{\tilde r} P_M=\frac{1}{\tilde r^3}\left[ (1-x)\lambda\int_0^1 
ds\;s^x\exp\left(-\frac{s+\lambda s^x}{\tilde r}\right)-\tilde r 
\exp\left(-\frac{1+\lambda}{\tilde r}\right)  \right] 
\label{derivate} 
\ee 
This clearly shows that when $x \geq 1$, $\partial_{\tilde r} P_M$ is 
negative for all the (positive) values of $\tilde r$, i.e. $P_M$ is a 
decreasing function of $\tilde r$. 
When $x<1$, the analysis is a bit more complicated. Let us show that 
$\partial_{\tilde r} P_M=0$ has at least one solution for $\tilde r > 
0$. First, when $\tilde r \to \infty$, from Eq. (\ref{derivate}) it is 
clear that $\partial_{\tilde r} P_M$ is negative. Second, the following inequality holds:  
\be 
\int_0^1 ds\; s^x 
\exp\left(-\frac{s+\lambda s^x}{\tilde r}\right)>\exp\left(-\frac{1+\lambda}{\tilde r}\right)\int_0^1 
ds\; s^x \sim \exp\left(-\frac{1+\lambda}{\tilde r}\right) 
\ee 
so that $\tilde r^3 \partial_{\tilde r} P_M$, and hence 
$\partial_{\tilde r} P_M$, is positive when $\tilde r \to 0$ (see 
Eq. (\ref{derivate})). Since $\partial_{\tilde r} P_M$ is a continuous 
function that is positive when $\tilde r \to 0$ and negative when 
$\tilde r \to \infty$, we conclude that $\partial_{\tilde r} P_M=0$ has at least one 
solution for $\tilde r \geq 0$.  We could rigorously prove that this 
solution is unique. However, for the sake of lightness, we present 
here a proof based on physical arguments. First of all, at large 
$\tilde r$, $P_M$ decreases when $\tilde r$ increases simply because the 
system does not have enough time to escape from ${\bf S}$ when the 
loading rate becomes too large. On the other hand, a decreasing $P_M$ when $\tilde r \to$ 0 
reflects the fact that at very large forces, the probability to fold 
into ${\bf N}$ is much higher than the probability to fold into ${\bf 
M}$, the probabilities being very low though. In this case, the more 
time spent at high force values, i.e. the lower $\tilde r$, the less 
probable to fold into ${\bf M}$. 

Because $P_M \to 0$ when both $\tilde r \to 0$ and $\tilde r \to 
\infty$, $P_M$ shows at least one maximum at intermediate values of 
$\tilde r$. Moreover, in the present case where the location of the 
kinetic barriers does not depend on the applied force, we find that 
there is a single maximum for $P_M$ when $x>1$.

\subsection*{Non-absorbing states: the quasi-static regime} 

In the more realistic case where the states are not absorbing, the 
dependence of $P_M$ with respect to $r$ has a different nature at low 
$r$. In this case fluctuations between ${\bf M}$ and ${\bf N}$ 
(passing through ${\bf S}$) tend to populate ${\bf N}$ at low forces. 
Indeed, by definition, the native state ${\bf N}$ is supposed to be 
much more stable than the other states of the system at zero force, 
namely ${\bf M}$ and ${\bf S}$.  Consequently, at low $r$ the system 
has enough time to populate the native state. Or in other words, 
$P_M(r)$ tends to its equilibrium value $\simeq\exp(-\Delta\Delta 
G_0/k_B T)$ when $r\to 0$.  In any case (for both $x\geq1$ and $x< 
1$), we hence expect that $P_M \to \exp(-\beta \Delta \Delta G_0) 
\approx 0$ when $r \to 0$ where $\Delta \Delta G_0 $ is the free 
energy difference between $M$ and $N$. 

\begin{figure}[t]
\centerline{\includegraphics[scale=0.45]{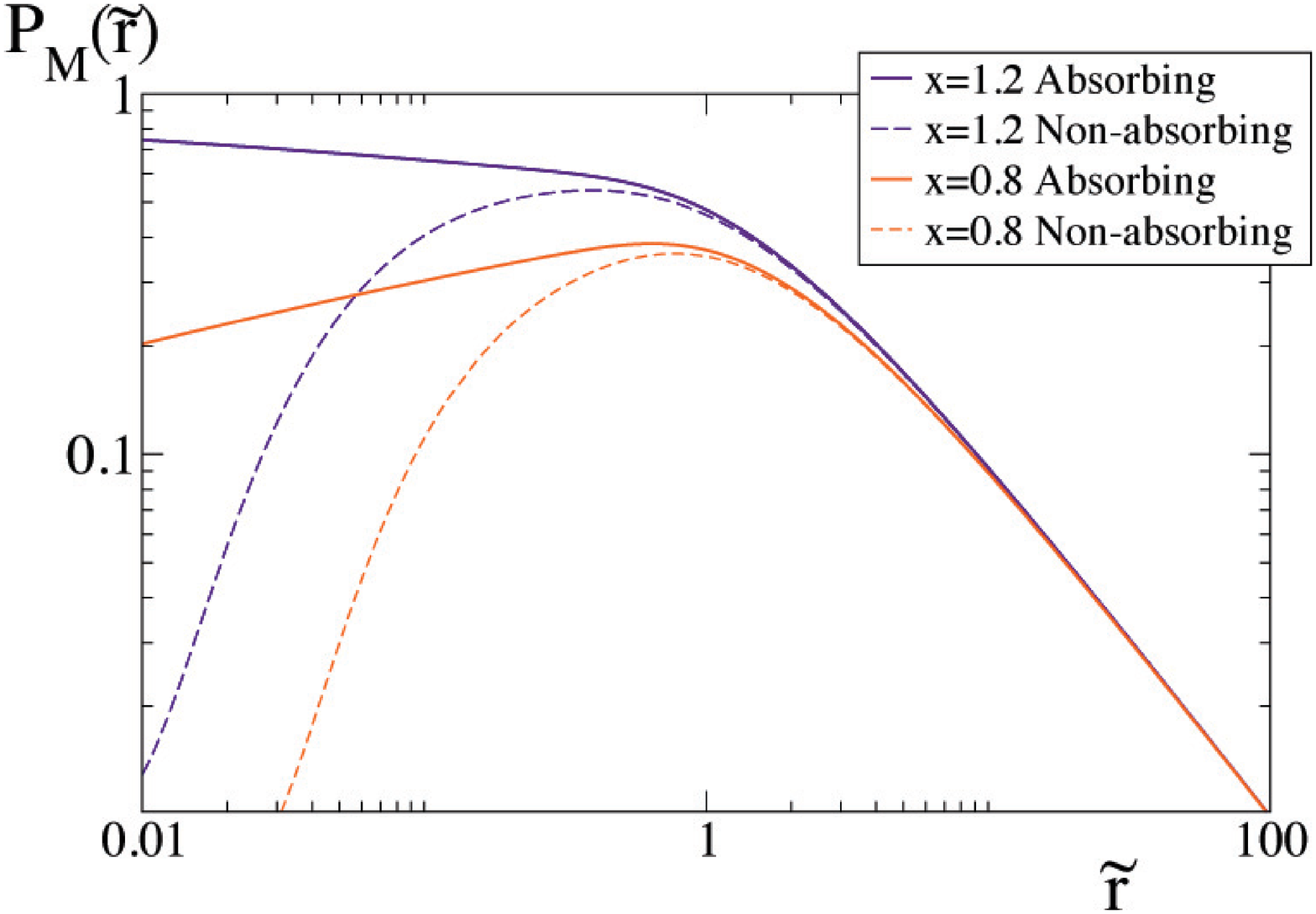}}
\caption{\label{f1S}}  
\end{figure}

To conclude, we can say that in a three-state system with 
force-independent location of the kinetic barriers, the misfolding 
probability $P_M$ shows always a bell-shape as shown in 
Fig. \ref{f1S}. However, the presence of the maximum may have a 
different cause depending on the value of the ratio $x=d_N/d_M$,
i.e. depending on the relative distances 
of the native and misfolded kinetic barriers to the stretched state.

\subsection*{Force-dependent location of the kinetic barriers} 

\begin{figure}[t]
\centerline{\includegraphics[scale=0.45]{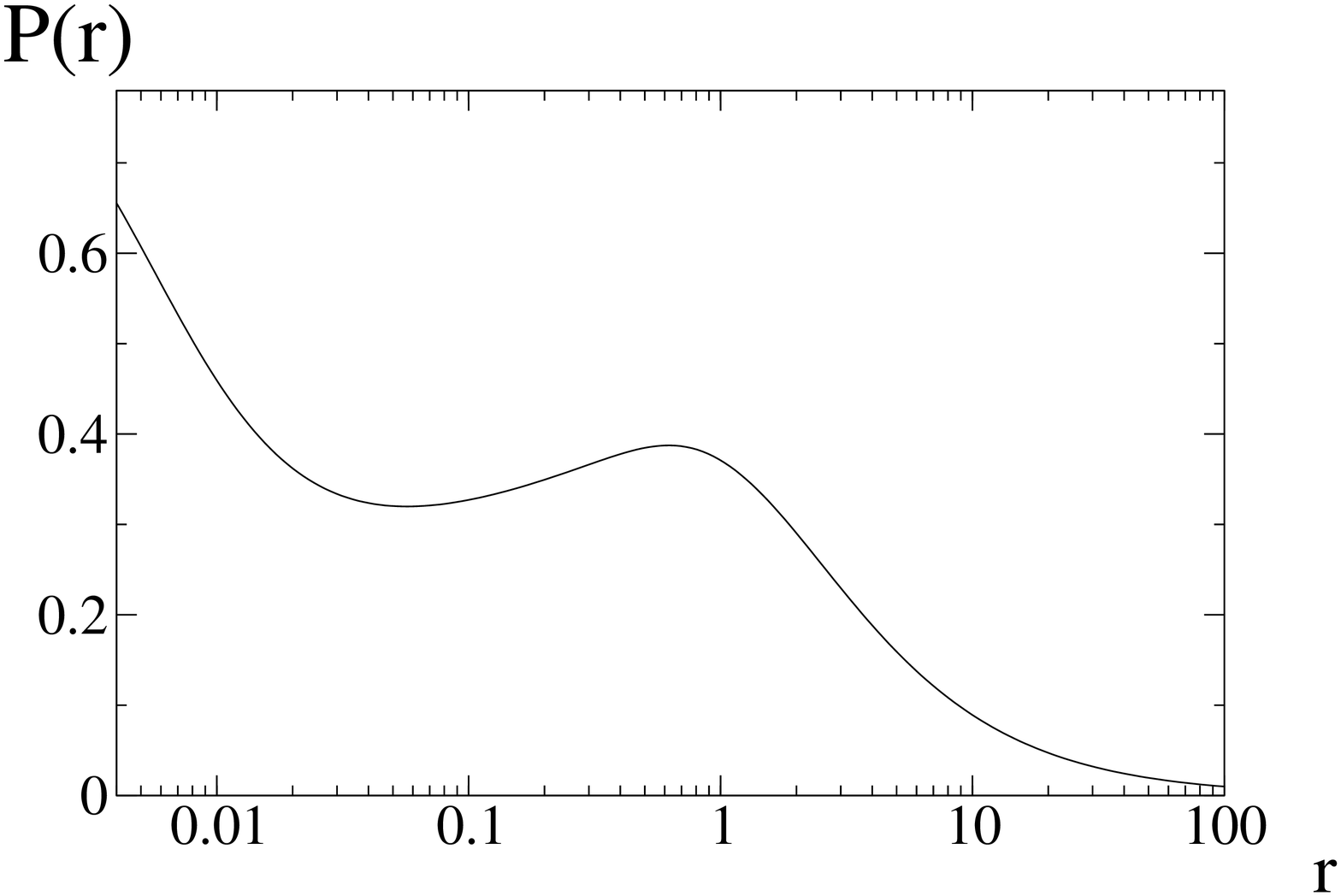}}
\caption{\label{f2S}}
\end{figure}

Numerical simulations in RNAS15 show a complex dependence of the misfolding
probability at the end of a force cycle with respect to the loading rate
$r$ (see Appendix \ref{app3} and Fig. \ref{f6}). This suggests
that RNAS15 cannot be modeled as a three-state model with force-independent
position of the kinetic barriers along the reaction
coordinate. Interestingly, in the three-state model described above,
still one can numerically study the effect of force-dependent positions
of the kinetic barriers on the shape of $P_M(r)$. Physically, a
dependence of $d_N$ and $d_M$ on the force corresponds to structural
changes in the corresponding transition states \cite{man1}. In the case of absorbing
states ${\bf N}$ and ${\bf M}$, and for a force protocol where the force
is released at constant rate $r$, the probabilities to be in the
different states ${\bf N}$, ${\bf M}$ and ${\bf S}$ at a given force $f$
read:

\bea  
\nonumber  
P_S(f)&=&\exp\left(-\frac{1}{r}\int_f^{\infty}dg\;\left(k_Ne^{-\beta d_N(g) g}+k_Me^{-\beta d_M(g) g}\right)\right)\\ 
\nonumber  
P_N(f)&=&\frac{1}{r}\int_f^{\infty}dg\; 
k_N\exp\left[-\beta d_N(g) g+\frac{1}{r}\int_g^{\infty}dh\;\left(k_Ne^{-\beta d_N(h) h}+k_Me^{-\beta d_M(h) h}\right)\right]\\  
P_M(f)&=&\frac{1}{r}\int_f^{\infty}dg\; 
k_M\exp\left[-\beta d_M(g) g+\frac{1}{r}\int_g^{\infty}dh\;\left(k_Ne^{-\beta d_N(h) h}+k_Me^{-\beta d_M(h) h}\right)\right] 
\label{ME5} 
\eea 
By playing with the force dependence of $d_N(f)$ and $d_M(f)$ we can 
obtain different shapes for the misfolding probability $P_M(f=0)$ that 
show several extrema along the $r$-axis. For instance, we can choose 
$d_M(f)<d_N(f)$ at low forces and $d_M(f)>d_N(f)$ at high forces. We 
then obtain a misfolding probability curve as the one shown in 
Fig. \ref{f2S}. The maximum at $r>0$ corresponds to a typical maximum 
of the force independent case $x=d_N/d_M < 1$, whereas the minimum at 
lower $r$ is due to a crossover from $x < 1$ to $x \geq 
1$. Interestingly, by solving the master equations \eq{s1} (see below) and by imposing the misfolded structure of RNAS15 to 
be an absorbing state, we obtain the same kind of dependence for the 
misfolding probability. This suggests that in RNAS15, $d_M(f) \leq d_N(f)$ 
at low forces. This also suggests that in the non-absorbing case, the 
low $r$-regime observed in the numerical simulations of RNAS15 is the 
consequence of a quasi-static regime that tends to populate the native 
state. 

\section{Appendix: Misfolding probability in RNAS15 \label{app3}}  

In RNAS15, we can estimate the misfolding probability by using the minimal 
structures model (MSM, see main text). Within this scheme, each 
configuration in the MSM can be labeled by $C_i$ where $i=1....N$, $N$ 
being the total number of MSM configurations. If $P_i(t)$ is the 
probability to be in the configuration $C_i$ at time $t$, the dynamics 
within the MSM is governed by the following set of master equations: 
\begin{eqnarray} 
\dot P_i(t)=-\sum_{\langle j 
\rangle}k_{i\rightarrow j}^f P_i(t)+ \sum_{\langle j 
\rangle}k_{j\rightarrow i}^f P_j(t)\quad \forall i \in [1;N]  
\label{s1} 
\end{eqnarray} 
where $\langle j \rangle$ counts for all the MSM configurations $C_j$ 
that are connected to $C_i$ {\it via} the sequential dynamics 
described in the Methods (see main text). $k_{i \to j}^f$ and $k_{j \to 
i}^f$ are the corresponding force-dependent closing/opening rates (see 
the Methods section). 

We numerically integrate this system by imposing a decreasing force at 
constant rate $r$ with the following initial 
condition: the molecule is in the stretched state 
($P_i(t=0)=1$ if $C_i={\bf S}$ and $P_i(t=0)=0$ otherwise) and the 
force $f=20$ pN. The curves we obtain are in good agreement with the 
experimental results (see Figs. \ref{f3S}).

Numerically, we have checked that our results remain unchanged using a 
coarse-grained description at the level of a few base-pairs in order 
to get results faster (simulations tend to be very slow when the number
of configurations starts to grow). In this case, we use the following two-state 
approximation. Let us suppose for instance that we coarse-grain the 
system of equations 
(\ref{s1}) at the level of $n_{bp}$ base-pairs (typically 
$n_{bp}=2,3$). If $k_{o,c}^*$ are the effective opening and closing 
rates, then $k_o^*+k_c^*=\lambda_s$ where $\lambda_s$ is the smallest 
eigenvalue of the $n_{bp}\times n_{bp}$ evolution matrix. The detailed 
balance condition imposes the value of the ratio $k_o^*/k_c^*$, hence it
determines the values of $k_o^*$ and $k_c^*$.

\newpage  

\section*{List of figures}

\subsection*{Figure \ref{f1}: Major and minor force-extension curves.}
(Color online) {\it Leftmost panel}: Optical tweezers experimental setup for single RNA 
manipulation (figure not to scale).  {\it Rightmost panel}: Experimental {\it major} and {\it minor} 
unfolding curves obtained from RNAS15 pulling experiments with optical 
tweezers\cite{Coll}.  The reported extension corresponds to the 
end-to-end distance of the RNA molecule plus the DNA/RNA hybrid 
handles. 

\subsection*{Figure \ref{f2}: Unfolding of the native structure.} 
(Color online) {\it Leftmost panel}: The RNAS15 
three-helix junction native structure composed of a stem $S$ (green) 
that branches into two hairpin loops $H_1$ (orange) and $H_2$(purple). 
Free energy of formation of the native state \cite{Vienna}: $\Delta 
G^{0}=-34.3$ kcal/mol $=-57$ $k_{B}T$ at room temperature ($298$K).  
{\it Rightmost panel}: Experimental {\it major}
unfolding curves compared with numerical results obtained from the 
sequential unfolding of the native structure (see 
text for details about the simulation procedure).

\subsection*{Figure \ref{f3}: Unfolding of the misfolded structure.} 
(Color online) (a): The most stable structure without stem (called, in this paper, the misfolded structure) is composed of two 
hairpins: $H_1^M$ (orange) and $H_2^M$ (red).  Its free energy of 
formation is equal to $\Delta G^{1}=-29$ kcal/mol $=-48.3$ 
$k_{B}T$.  (b): Experimental {\it minor}
unfolding curves compared with numerical results obtained from the 
sequential unfolding of the misfolded structure on the left. (c): Curves obtained from sequential simulations (see text) of the unfolding of the
individual hairpins $H_1^M$ and $H_2^M$ that compose the misfolded 
structure. Continuous lines represent a low bandwith average of the 
force-extension data.

\subsection*{Figure \ref{f4}: The minimal structures model (MSM).} 
(Color online) {\it Upper panel}: Schematic representation of the  
sequential model for multi-hairpin structures. The only allowed  
transitions are the opening and closing of the base pairs located at  
the base of the hairpins (shown as thick bonds) where the  
force is applied. {\it Lower panel}: How to build the ensemble of  
configurations of the MSM.  The intermediate configurations resulting  
from the sequential unfolding of either $N$ or $M$ are composed of  
hairpins and regions of unpaired bases (shown in blue). Then, the final MSM  
ensemble results from the combination of all the different hairpins  
and unpaired regions. In the example shown here, two hairpins ($A$ and  
$B$) are combined together to form a configuration where the two  
original hairpins are separated by a region of unpaired bases.

\subsection*{Figure \ref{fbox2}: Free energy corrections.} 
(Color online) Upper panel (a):  schematic representation of a generic configuration $C$ of the  
MSM. It is composed of hairpins and regions of unpaired  
bases. The free energy correction of a given configuration $C$ at  
force $f$, $G_c(C,f)$, is given by the sum of the independent free energy  
contributions coming from all different unpaired regions. Lower  
panel: function $F(n,f)$, defined as the free energy of an $n$-bases  
polynucleotide chain minus the mechanical free energy of the  
fully extended chain averaged over all possible segments of that  
length $n$ along the RNAS15 sequence. We find that $F(n,f)$ is  
approximately linear with $n$, $F(n,f)\approx a_f (n-5)$. The  
coefficient $a_f$ as a function of the force is plotted in the inset  
of the figure.

\subsection*{Figure \ref{f5}: Dynamic force spectroscopy results.}  
(Color online) Experimental results compared to numerical  
simulations in the MSM. The MSM parameters are: $A=0.3$ $k_B T$ and  
$k_a=10^7 {\rm s}^{-1}$.  (a): Unfolding and refolding major curves at  
loading rate $r \simeq 20 {\rm pN.s^{-1}}$.  (b): Distribution of  
breakage forces, i.e. the force at which the molecule unfolds,  
obtained from the major unfolding curves at $r \simeq 20 {\rm pN.s^{-1}}$  
(distributions have been obtained from 900 (2000) trajectories in the  
experiments (simulations)) and $r \simeq 12 {\rm pN.s^{-1}}$  
(distributions have been obtained from 400 (2000)  
trajectories in the experiments (simulations)).

\subsection*{Figure \ref{f6}: Misfolding probability and three-state model.}  
(Color online) Upper panel: Representation of the three-state model including the  
stretched, native and misfolded states. The misfolded state acts as a  
kinetic trap for the folding transition between the stretched state and  
the native state. Lower panel: Misfolding probability (computed at the  
end of the relaxing process) as a function of the unloading rate. The  
experimental points correspond to $r= 20 {\rm pN.s^{-1}}$ and $r= 12  
{\rm pN.s^{-1}}$.

\subsection*{Figure \ref{ffree}:  (Appendix A)}  
(Color online) Free energy as a function of the number of
opened base pairs for the two hairpins forming the $M$ structure,
$H_{1}^{M}$ (red) and $H_{2}^{M}$ (blue), at the critical force where
both the folded and the unfolded hairpins are equally stable (critical
force values are around 10 and 11 pN for $H_{1}^{M}$ and $H_{2}^{M}$
respectively). Results shown are obtained by using the Vienna package
\cite{Vienna}.

\subsection*{Figure \ref{MNS}: (Appendix B)} 
Three-state model with three states
${\bf N},{\bf M},{\bf S}$. ${\bf S}$ is an intermediate state
on-pathway from the misfolded to the native state.  The four possible
rates for $k_{a\to b}^f$ are also shown.

\subsection*{Figure \ref{f1S}:  (Appendix B)} 
(Color online) Misfolding probability $P_M$ as a function of the adimensional rate $\tilde r$ for the three-state model with
force-independent positions of the kinetic barriers and
native/misfolded absorbing states. The full curves have been obtained
by numerically integrating Eq. (\ref{PM}) with $\lambda=1/x$ so that
$k_N=k_M$. The dashed curves show the corresponding case where the
native/misfolded states are non absorbing. In this case, we denote by
$d_N^{\dag}$, respectively $d_M^{\dag}$, the distance from states
${\bf N}$, ${\bf M}$ respectively, to the position along the reaction
coordinate of the kinetic barrier separating these states from ${\bf
S}$. The curves have been obtained using $k_{N \to S}=k_N
\exp(-\beta(\Delta G_0+fd_N^{\dag}))$ and $k_{M \to S}=k_M
\exp(-\beta(\Delta G_1+fd_M^{\dag}))$ with
$k_N=k_M=\beta=d_N^{\dag}=d_M^{\dag}=1$. $\Delta G_0=20$ and $\Delta
G_1=10$ correspond to the free energy of formation of the native and
misfolded states, respectively, at zero force.

\subsection*{Figure \ref{f2S}:  (Appendix B)} 
Misfolding probability $P_M$ as a function of the
rate $r$ in the case of a force-dependent position of the barrier
between the native and the stretched state in the three-state model
with absorbing native/misfolded states. The curve has been obtained by
taking $d_N=0.8d_M$ for $f>5.5$ pN and $d_N=1.25d_M$ for $f\leq 5.5$
pN. We have also used $k_N=k_M=\beta=1$.

\begin{figure}[t]
\centerline{\includegraphics[scale=0.5]{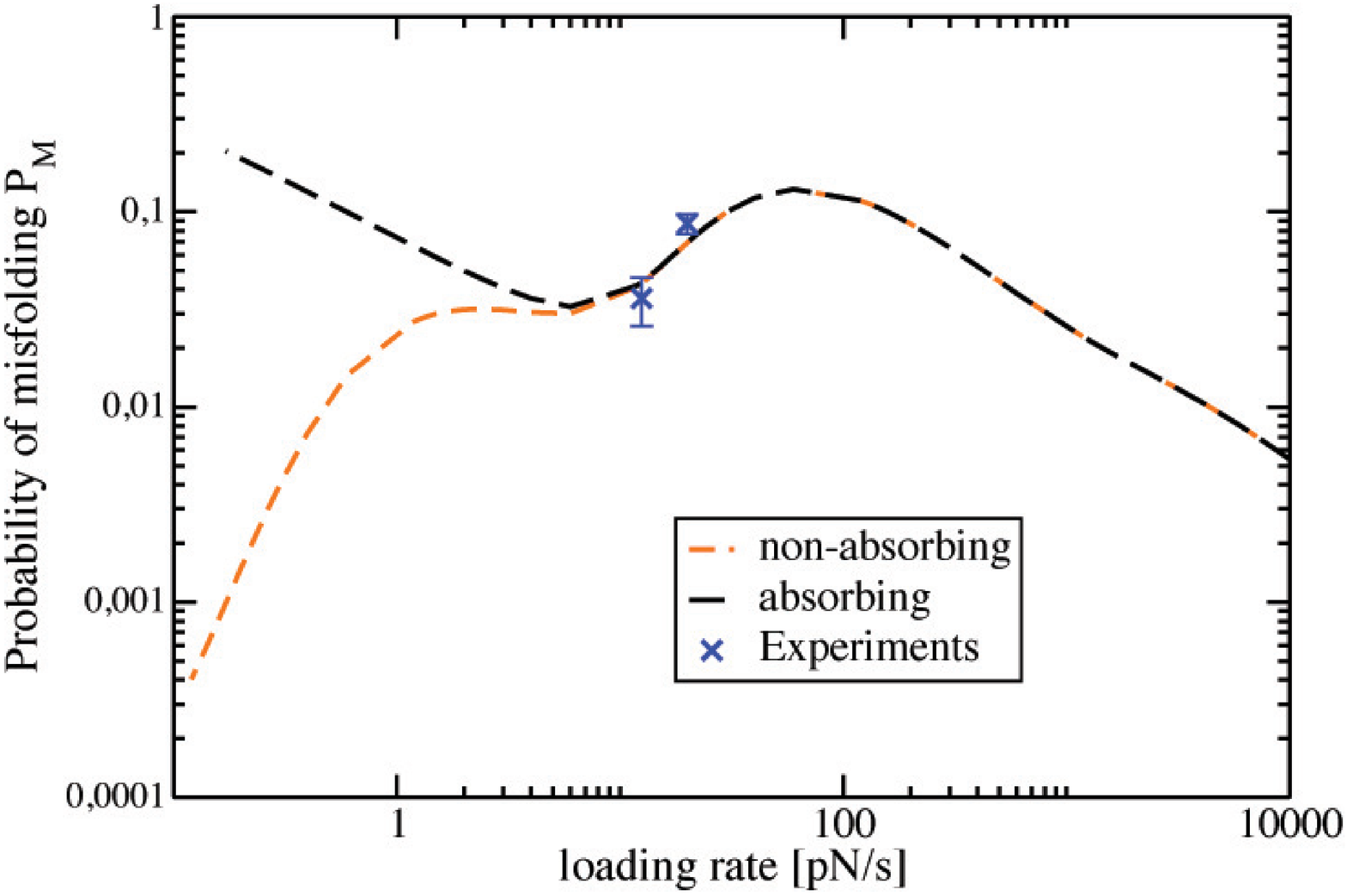}}
\caption{\label{f3S}}  
\end{figure}

\subsection*{Figure \ref{f3S}:  (Appendix C)} 
(Color online) Misfolding probability obtained from the set of
master equations \eq{s1} describing the folding kinetics in the MSM.
The dashed black lines correspond to the case where the misfolded
state is absorbing. 

\newpage

\end{document}